\documentclass[11pt,preprint]{aastex}

\usepackage{graphicx,amsmath,booktabs,threeparttable,url}
\usepackage{hyperref}

\usepackage{threeparttable}

\usepackage[usenames,dvips]{color}

\usepackage[normalem]{ulem}

\usepackage{natbib}
\begin{document}

\title{IMPACT OF TYPE Ia SUPERNOVA EJECTA ON BINARY COMPANIONS IN THE SINGLE-DEGENERATE SCENARIO }

\author{Kuo-Chuan Pan$^1$, Paul M. Ricker$^1$ and Ronald E. Taam$^{2,3}$}
\affil{$^1$Department of Astronomy, University of Illinois at Urbana-Champaign, 1002 W. Green Street, Urbana, IL 61801; kpan2@illinois.edu, pmricker@illinois.edu}
\affil{$^2$Department of Physics and Astronomy, Northwestern University, 2145 Sheridan Road, Evanston, IL 60208; r-taam@northwestern.edu}
\affil{$^3$Academia Sinica Institute of Astronomy and Astrophysics, P.O. Box 23-141, Taipei 10617, Taiwan}

\begin{abstract}

Type Ia supernovae are thought to be caused by thermonuclear explosions of a carbon-oxygen white dwarf 
in close binary systems. In the single-degenerate scenario (SDS), the companion star is non-degenerate 
and can be significantly affected by the explosion. We explore this interaction by means of 
multi-dimensional adaptive mesh refinement simulations using the FLASH code. We consider several 
different companion types, including main-sequence-like stars (MS), red giants (RG), and helium stars 
(He). In addition, we include the symmetry-breaking effects of orbital motion, rotation of the 
non-degenerate star, and Roche-lobe overflow. A detailed study of a sub-grid model for Type Ia 
supernovae is also presented. 
We find that the dependence of the unbound stellar mass and kick velocity on the initial binary 
separation can be fitted by power-law relations. By using the tracer particles in FLASH, the 
process leading to the unbinding of matter is dominated by ablation, which has usually been neglected 
in past analytical studies. The level of Ni/Fe contamination of the companion that results from the 
passage of the supernova ejecta is found to be a $\sim 10^{-5} M_{\odot}$ for the MS star, 
$\sim 10^{-4} M_{\odot}$ for the He star, and $\sim 10^{-8} M_{\odot}$ for the RG. 
The spinning MS companion star loses about half of its initial angular momentum during the impact, 
causing the rotational velocity to drop to a quarter of the original rotational velocity,
suggesting that the Tycho G star is a promising progenitor candidate in the SDS.

\end{abstract}



\section{INTRODUCTION}
Type Ia supernovae (SNe~Ia) are considered to be ``standardizable'' candles when measuring the distance of distant objects \citep{1993ApJ...413L.105P} and thus play an important role in cosmology, helping determine key cosmological parameters \citep{1992ARA&A..30..359B, 1998AJ....116.1009R, 1998ApJ...507...46S, 1999ApJ...517..565P}. 
Most cosmological applications using SNe~Ia are based on an assumption that the homogeneity of SNe~Ia applies to high-redshift SNe~Ia.
However, this assumption is questionable until we fully understand the nature of the progenitors of SNe~Ia.
According to recent studies, SNe~Ia are believed to be thermonuclear explosions of carbon-oxygen (CO) white dwarfs (WDs), but the ignition mechanism and the progenitor systems are still unknown. 

Current popular progenitor models are classified into two major scenarios, the single-degenerate scenario (SDS, \cite{1973ApJ...186.1007W, 1982ApJ...257..780N}) and the double-degenerate scenario (DDS, \cite{1984ApJS...54..335I, 1984ApJ...277..355W}).
The SDS involves a CO WD accreting matter from a non-degenerate binary companion, such as main-sequence (MS) stars, red giants (RG), or helium stars (He). 
The CO WD eventually becomes unstable and then, when the mass reaches the Chandrasekhar limit ($M\sim 1.4 M_\odot$), it explodes as a SN~Ia. 
The DDS instead involves two CO WDs, with total mass larger than the Chandrasekhar mass, that merge together because of the emission of gravitational waves.  

With the DDS, the lack of hydrogen spectral lines is easy to explain; however, the homogeneity of the explosion energy is difficult to account for since the total mass of the binary system varies from case to case. 
Furthermore, it is possible that WD-WD mergers end with the production of an ONeMg WD followed by 
accretion-induced collapse to a neutron star \citep{1985ApJ...297..531N}.  \cite{2012arXiv1202.5472B} 
calculate the merger rate of binary WDs in the Galactic disk based on the observational data 
in the Sloan Digital Sky Survey. They conclude that the merger rate of binary WDs with 
super-Chandrasekhar masses would not significantly contribute to the SNe~Ia rate.
The SDS produces a relatively fixed explosion energy, which accounts for the homogeneity. 
The non-degenerate companions in the SDS, however, are usually hydrogen-rich which leads to the H/He contamination problem in the SDS (see \cite{2000tias.conf...33L} and \cite{2000ARA&A..38..191H} for detailed descriptions of the advantages and weaknesses of the SDS and DDS).  

Recent studies based on the delay time distribution (DTD) suggest the need for at least a two-component model for the DTD. 
In particular, \cite{2005ApJ...629L..85S}, \cite{2006MNRAS.370..773M} and \cite{2010AJ....140..804B} found that the observations can be fitted with a short-delay-time population ($\sim 10^8$~yr) and a long-delay-time population ($3-4$~Gyr).

The calculations of SN~Ia rates and DTD using binary population synthesis can be matched with both the SDS \citep{2003IAUS..214..109H, 2008ApJ...679.1390H} and the DDS \citep{2008ApJ...683L..25P, 2009ApJ...699.2026R}.
The uncertainties in the observed DTD, however, are dominated by the uncertainties in galactic stellar populations and  star formation histories \citep{2011arXiv1111.4492M}.
Thus, it is still hard to distinguish between the viability of the SDS and the DDS progenitor models without more detailed observations.


With detailed binary calculations, the long-delay-time population can be understood in terms of progenitor systems characterized by a MS-like companion in the MS-WD channel \citep{2004ApJ...601.1058I,2010MNRAS.401.2729W, 2010MNRAS.404L..84W} and/or by an RG in the RG-WD channel \citep{1999ApJ...522..487H,  2008ApJ...679.1390H}. 
In contrast, the short-delay-time population may consist of systems with a massive MS star in the MS-WD channel \citep{2008ApJ...679.1390H} or with a He-star in the He-WD channel \citep{2009MNRAS.395..847W}.
Therefore, the more likely explanation is that both SDS and DDS have contributed to SNe~Ia.

Several detailed hydrodynamics simulations with the SDS have been studied in the past decade.
\cite{2000ApJS..128..615M} examined the impact of a SN~Ia on a MS star, a sub-giant, and a RG using two-dimensional Eulerian hydrodynamics. 
\cite{2000ApJS..128..615M} found that the MS star and sub-giant companions lose $15\%$ of their mass after the explosion, and the RG companion loses about $96\%-98\%$ of its envelope.
They also found that the impact of SN ejecta with the companion star creates a hole with 
an opening angle of $\sim 30^\circ$ in the high-velocity ejecta and $\sim 40^\circ$ in the low-velocity ejecta. 
This hole corresponds to $7\%-12\%$ of the ejecta's surface, making the supernova remnant (SNR) anisotropic and 
potentially affecting Si~II spectral line shapes. 

\cite{2008A&A...489..943P} re-examined the MS-WD channel and updated these results 
by considering the pre-supernova binary evolution from \cite{2004ApJ...601.1058I} and
using three-dimensional smoothed particle hydrodynamics (SPH) simulations. 
They found strong dependences of unbound mass and kick velocity on initial binary separation with power-law relations.
Because of the relatively small number of particles in their SPH simulations, 
they were unable to reproduce the turbulence in \cite{2000ApJS..128..615M}.
However, their comparison to \cite{2000ApJS..128..615M} found that the unbound mass does not depend much on the numerical method used. 
Although the unbound mass is not sensitive to the turbulence around the MS star, other important physical quantities, e.g. contaminated SN ejecta on the companion star, may be sensitive to small-scale turbulence. 

In contrast with \cite{2000ApJS..128..615M}, \cite{2010ApJ...715...78P} studied the He-WD channel from \cite{2009MNRAS.395..847W} for the short-delay-time population, using two-dimensional Eulerian hydrodynamics simulations with adaptive mesh refinement. 
They also found the same power-law relations in unbound mass and kick velocity as \cite{2008A&A...489..943P}, but with different power-law indices. 
An upper limit of the ratio of Ni/Fe contamination to the helium abundance was found to be $\sim 9-50 \times 10^{-4}$, which is higher than the solar abundance ratio. 
This result suggests the possibility of detecting remnant helium stars in their SNR.

In this paper, we examine the SDS using the above discussed companion star models to determine whether the contamination problem of H/He in the companion envelope can be overcome.
We revisit this problem using modern adaptive mesh refinement (AMR)  
and different companion models that incorporate the effects of binary evolution.
We include the symmetry-breaking effects of orbital motion, rotation of the non-degenerate star, and Roche-lobe overflow (RLOF).
In the next section, the numerical methods and the construction of initial setups are described.
Our numerical results for different companion models are reported in Section 3. 
In Section 4 we present a parameter study made by varying significant physical quantities and then discuss the possible observational implications. 
In the final section, we summarize our results and conclude.


\section{NUMERICAL METHODS AND MODELS}

\subsection{Numerical Codes}

Two different codes are employed in this work. 
The first one is a stellar evolution code used to construct 
the non-degenerate stellar models.
The second code is a multi-dimensional hydrodynamics code to simulate the impact of type~Ia supernova ejecta. 

The stellar evolution code used is MESA\footnote{\url{http://mesa.sourceforge.net/index.html}} (Modules for Experiments in Stellar Astrophysics,  \cite{2011ApJS..192....3P}). 
MESA combines many sophisticated numerical methods to control time and space resolutions, and many state-of-the-art physical modules to solve the stellar structure. 
For a given initial stellar mass and metallicity, MESA evolves stars to a certain age and provides all the physical quantities we need for the hydrodynamics simulations.

For the SN~Ia simulations, we use FLASH\footnote{\url{http://flash.uchicago.edu}} version~3 \citep{2000ApJS..131..273F, 2008PhST..132a4046D}.
FLASH is a parallel, multi-dimensional hydrodynamics code based on block-structured adaptive mesh refinement (AMR).
To solve the Euler equations, we use the split piecewise parabolic method (PPM) solver \citep{1984JCoPh..54..174C} in FLASH.
The equation of state (EOS) applied is the Helmholtz EOS \citep{2000ApJS..126..501T}, which is interpolated from a precomputed table of the Helmholtz free energy. It includes contributions from radiation, completely ionized nuclei, and degenerate electrons and positrons for an optically thick mixture of gas and radiation in local thermodynamic equilibrium. 
Magnetic fields are ignored, but self-gravity is considered and solved using the multipole Poisson solver in 2D and using the multigrid Poisson Solver in 3D \citep{2008ApJS..176..293R}. 
Particle modules in FLASH~3 have the ability to treat both active and passive particles.
Active particles are massive particles that interact through 
gravity with other active particles and fluid elements.
Passive particles are massless particles that only follow the motion of fluid elements in Lagrangian coordinates.
We use active particles to represent the CO WD and the core of the RG companion and passive particles to monitor the motion of fluids.

\subsection{Non-Degenerate Companion Models}

\begin{table}
\begin{center}
\caption{The non-degenerate companion models. \label{tab1}}
\begin{tabular}{lcccccc}
\\
\tableline
	   & $M$ \tablenotemark{a}              & $R_*$ \tablenotemark{b}& $\rho_c$ \tablenotemark{c}& $P_c$\tablenotemark{d}& $T_c$ \tablenotemark{e}& $t_{dyn}$\tablenotemark{f}\\
Model & (M$_{\odot}$) & (km) & (g/cm$^3$) & (dyne/cm$^2$) & (K)   & (sec) \\
\tableline
HCV' & 1.027 & $6.76 \times 10^{5}$ & $112.3$ & $1.90 \times 10^{17}$ & $1.48 \times 10^{7}$ & $1.54\times 10^3$  \\
MS-WD & 1.17 & $5.51\times 10^{5}$ & $68.9$ & $1.33 \times 10^{17}$ & $1.44 \times 10^{7}$ & $1.06\times 10^{3}$ \\
RG-WD & 0.98 & $2.15\times 10^{7}$ & $3.17\times10^{-3}$ \tablenotemark{\dagger} & $5.50\times 10^{11}$\tablenotemark{\dagger} & $1.88\times 10^{6}$\tablenotemark{\dagger} & $2.82\times 10^{5}$ \\
He-WDc & 1.007 & $1.36\times 10^{5}$ & $6.10 \times 10^3$ & $5.12 \times 10^{19}$ & $1.28 \times 10^{8}$ &  $1.40\times 10^{2}$ \\
\tableline
\end{tabular}
\tablenotetext{a}{Stellar mass, $M$}
\tablenotetext{b}{Stellar radius, $R_*$}
\tablenotetext{c}{Central density,$\rho_c$}
\tablenotetext{d}{Central pressure, $P_c$}
\tablenotetext{e}{Central Temperature, $T_c$}
\tablenotetext{f}{The dynamical time scale, $t_{dyn}=1/2(G\rho)^{-1/2}$}
\tablenotetext{\dagger}{For the RG-WD model, the central density, pressure and temperature are the central values of the envelope mapped in FLASH, not the core values in MESA.}
\end{center}
\end{table}
\begin{figure}
\plotone{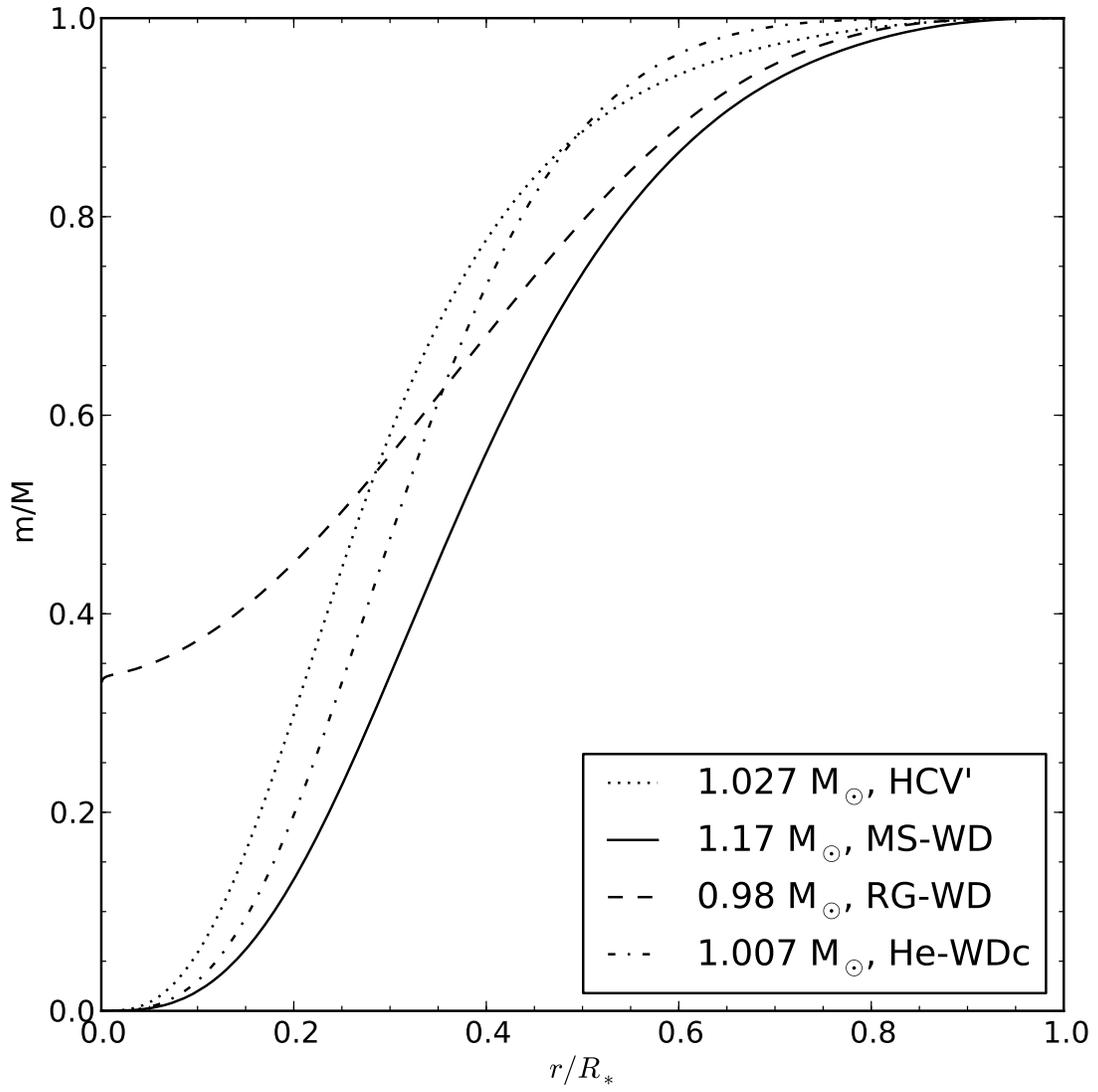}
\caption{\label{models} Mass vs. radius profiles for the main sequence, red giant, and helium star models. For the RG-WD scenario, only the envelope mass of the RG is plotted here.}
\end{figure}

In order to compare our setup with previous work by \cite{2000ApJS..128..615M} and \cite{2008A&A...489..943P}, we use MESA to create a MS star companion model similar to the HCV scenario in \cite{2000ApJS..128..615M}.
The HCV scenario in \cite{2000ApJS..128..615M} is a hydrogen cataclysmic variable system consisting of a MS star and a CO~WD. 
We assume the binary system is in RLOF and has a binary separation $a=3R_*$ at the onset of the SN~Ia explosion.
The MS star companion in the HCV scenario has a mass $M=1.017$~M$_\odot$ and radius $R_*=6.8\times 10^5$ km.
Because of the lack of certain physical data in the HCV scenario, we created a similar MS star model, named HCV', with $M=1.027$~M$_\odot$, and $R_*=6.76 \times 10^5$km.

For more realistic models, we use models of SN~Ia progenitor candidates taken from detailed binary calculations. 
Based on current studies, we chose three possible stellar types of binary companions in the SDS: MS stars, RGs, and He stars.

\cite{2004ApJ...601.1058I} studied the evolution of binaries consisting of evolved MS stars with WDs. 
They also investigated possible channels that may eventually evolve to SNe~Ia. 
We chose parameters from the case with final mass $M_{{\rm d;f}}=1.17 M_\sun$ for the MS star and initial orbital period $P=1$~day to construct a model of a MS-WD binary system (the system is denoted by MS-WD in our simulations).

\cite{1999ApJ...522..487H} have proposed an alternative binary channel consisting of a WD and a low-mass RG.
The accreting WD in this model has a strong optically thick wind, which broadens the range of parameters leading to SNe~Ia.
We also adopt the case P1 in \cite{1999ApJ...522..487H} for our simulations (denoted by RG-WD here).
This system consists of a one solar mass WD and a two solar mass RG with an  initial orbital period $P_{\rm 0}=300$~days. 
The system ends with a SN~Ia after $7.2\times 10^5$~years; the mass of RG becomes $0.98 M_\sun$ at the onset of the SN~Ia explosion.

For the helium star model, 
we rely on the model originally proposed by \cite{2009MNRAS.395..847W} as well as the case where the binary system is in a stable He-shell burning phase at the onset of the SN explosion (Case 2 in \cite{2009MNRAS.395..847W}). 
This model also is the He-WDc model applied in \cite{2010ApJ...715...78P}, but with nonuniform chemical composition. 
The mass of this helium star model is $1.007 M_\sun$ (denoted by He-WDc).

In MESA, we do not follow full binary evolution but use the initial conditions with a constant mass-loss rate, estimated from the above references as an approximation. 
The mass-loss rate used is taken to be the ratio of the mass change to the evolution time in the above references.
All models we created from MESA are summarized in Table~\ref{tab1}, and the mass versus radius profiles are shown in Figure~\ref{models}.

\subsection{Initial Setup}

We conduct two- and three-dimensional hydrodynamics simulations 
to investigate the impact of SN~Ia ejecta on the binary companions.  
In the three-dimensional calculations, 
we rely on Cartesian coordinates; the simulation box dimensions are set to a size equal to $30$ times the radius of the non-degenerate companion star ($R_*$) in all directions.
The companion star is located at the center of the simulation box for convenience. 
Two-dimensional models are calculated in cylindrical coordinates due to the axial symmetry if the orbital motion of the binary system is ignored. 
The simulation box dimensions are set to $15 R_*$ in the radial ($r$) direction and $30 R_*$ in the axial ($z$) direction. 
We interpolated the one-dimensional model onto the FLASH grids using up to 10 levels of refinement based on the magnitudes of the second derivatives of gas density and pressure.  
Each AMR block contains $8^3$ zones in the three-dimensional box and $8\times 16$ zones in the two-dimensional box. 
This corresponds to an effective uniform resolution of $4096^3$ in three-dimension and $4096\times 8192$ in two dimensions. 
To simplify the problem, the compositions from MESA are adjusted to hydrogen ($^1$H), helium ($^4$He) and carbon ($^{12}$C) only (companion material).
We use outflow boundary conditions for fluids and isolated boundary conditions for the Poisson solver.

In an Eulerian hydrodynamics code like FLASH, physical properties are calculated at fixed spatial positions. 
It is difficult to trace fluid elements in a time sequence.
However, FLASH has the ability to trace fluid elements using passive particles.
Passive particles are massless particles that only follow the motion of fluid elements in Lagrangian coordinates without interacting with fluids.
In our simulations, $10^6$ passive particles are distributed with the gas density in order to study the history of shocked gas. 

For the RG model, although we have AMR, the core of the RG still cannot be resolved with current computational resources. 
Thus, we use a rigid spherical particle cloud with $3 \times 10^5$ active particles to represent it. 
We artificially modify the physical quantities within the core region of the RG to increase linearly with  the radius and then reconstruct the model on multi-dimensional grids in FLASH.
Therefore, the mass difference between the real stellar model and the reconstructed model in FLASH is the core mass of the RG ($M_{\rm RG,c}=0.311 M_\odot$). 
The radius of the particle cloud is set to three times the smallest zone spacing; the force on the cloud uses cloud in cell (CIC) interpolation.
We utilize a particle cloud, instead of a single particle, to avoid the problem of force anisotropy in CIC \citep{2008ApJ...672L..41R}.

In order to validate our models in hydrostatic equilibrium and to reduce the geometrical 
distortion between one dimension and multi-dimensions, all models are relaxed on the multi-dimensional Eulerian grid by artificially damping the gas velocity for 5 dynamical time scales ($t_{\rm dyn}=1/2(G\rho)^{-1/2}$). 
During this time, a damping factor, smoothly increased from 0.7 to 0.99 on the gas velocity, is imposed at each timestep.
We ensure that the companion models remain relatively stable during the relaxation and that the Mach number is always smaller than $0.01$. 

After relaxation, we introduce asymmetric effects from orbital motion by adding a WD into the simulation box (three-dimensional runs only).
The WD is represented by another particle cloud with $3\times 10^5$ particles and its mass is set to $M=1.378 M_\odot$ (we assume the binary system is about to explode).
In the three-dimensional simulations, the orbital plane is set on the $x-y$ plane and the WD is placed on the positive $x-$axis (positive radial direction in two dimension) with a binary separation $a$.
Positive $z-$axis is set to the direction of angular momentum. 
The binary system is assumed to be in RLOF and the binary separation is calculated using equation~\ref{RLOF} \citep{1983ApJ...268..368E}: 
\begin{equation}
\frac{R_L}{a}=\frac{0.49 q^{2/3}}{0.6q^{2/3}+ \ln(1+q^{1/3})}, \label{RLOF}
\end{equation}
where $R_L$ is the radius of the Roche lobe-filling star, and $q$ is the mass ratio.
The non-degenerate binary companion has a spin with a spin-to-orbit ratio of $0.95$. 
In this phase, the maximum AMR level is reduced to 8 levels to save computation time and the 
AMR level is forced to the maximum level (8) in two spherical regions centered on the WD and the companion star within a radius of $R < 1.8 R_*$. 
AMR levels in other regions are calculated from the second derivative of gas density and pressure 
and limited to 6 levels (a situation denoted by $6/8$ levels, equivalent to a $1024^3$ uniform grid in three dimensions and $1024 \times 2048$ in two dimensions). 

\subsection{The Supernova Model \label{sec_snmodel}} 

After the binary system is placed in the simulation box, 
we simulate the binary evolution for a few orbital periods. 
During this phase, mass transfer via RLOF occurs and the companion 
star is slightly distorted due to the tidal force. 
With current computational resources, we cannot simulate the whole delay time in multi-dimensional hydrodynamics simulations. 
We therefore simulate the binary evolution for only a few orbital periods ($< 3 P_{\rm orb}$) and assume the morphology does not change much once the RLOF is stable.  
We tested the impact of a SN~Ia explosion on the binary companion using different binary evolution times. 
We find that binary evolution only affects the morphology of the SNR at a late time and does not alter significantly the impact on the companion star.
Thus, all the runs we describe later ignore the binary evolution.
The orbital velocity and spin, however, are still considered. 
 
To introduce a SN~Ia explosion, we remove the WD particle cloud and replace it with a high-density and high-temperature gas. 
The physical quantities used to describe the SN~Ia explosion are taken from the W7 model by \cite{1984ApJ...286..644N}. 
It has a mass $M_{\rm sn}=1.378 M_\odot$, total explosion energy $E_{\rm sn}=1.233 \times 10^{51}$ erg, and average ejecta speed $v_{\rm sn}=8.527 \times 10^3$ km~s$^{-1}$. 
The kinetic energy of the WD from the orbital motion also is added to the SN.
We rely on nickel ($^{56}$Ni) as a tracer for the SN~Ia material. 

Although we have AMR, the WD is still much smaller than the minimum zone spacing.
Thus, a sub-grid model is required to approximate the SN~Ia explosion. 
To minimize grid effects, we use a small spherical region with a radius equal to fifteen times the smallest zone spacing to represent the SN material.
Even if we put in the same amount of explosion mass and energy, the results are actually sensitive to the initial distribution of the SN ejecta within this small region.
The detailed setup of the sub-grid SN model and the corresponding behavior will be discussed in the next section. 


\section{RESULTS}

\begin{table}
\begin{center}
\caption{Simulations \label{tabAllruns}}
\begin{tabular}{lcccccc}
\\
\tableline
          & $a$ \tablenotemark{a} & $\delta M_{ub} \tablenotemark{b}$    & $\delta v_{kick}$  \tablenotemark{c}& $v_{orb}$ \tablenotemark{d}& $M_{str}/M_{abl}$ \tablenotemark{e}\\
Case & (cm) & ($M_\odot$) & (km/sec)      & (km/sec) & ($M_\odot$) \\
\tableline
MS-r  & $ 1.51 \times 10^{11}$& $2.08\times 10^{-1}$  & 114& 256 & 0.65\\
MS-4 & $ 2.20\times 10^{11}$ & $5.00\times 10^{-2}$ & 58.8 & 212 & 0.43\\
MS-5 & $ 2.75\times 10^{11}$ & $2.41\times 10^{-2}$ & 39.6 & 190 & 0.73\\
MS-Nr & $ 1.51\times 10^{11}$& $1.80\times 10^{-1}$ & 112 & 0 & 0.49\\
MS-N4 & $ 2.20\times 10^{11}$& $4.35\times 10^{-2}$ & 57 & 0 & 0.39\\
MS-N5 & $ 2.75\times 10^{11}$& $2.31\times 10^{-2}$& 41 & 0 & 0.58\\
MS-2D-Nr  & $ 1.51 \times 10^{11}$& $1.84\times 10^{-1}$  & 113 & $-$ & $-$ \\
MS-2D-N4  & $ 2.20 \times 10^{11}$& $4.21\times 10^{-2}$  & 59.5 & $-$ &$-$ \\
MS-2D-N5 & $ 2.75 \times 10^{11}$& $2.08\times 10^{-2}$  & 36 & $-$ &$-$ \\
RG-r & $ 6.13\times 10^{12}$& $6.37\times 10^{-1}$ & 12 & 41.8 & 0.01\\
RG-4 & $ 8.58\times 10^{12}$& $5.93\times 10^{-1}$& 25 & 35.3 & 0.06\\
RG-5 & $ 1.07\times 10^{13}$& $5.07\times 10^{-1}$& 19 & 31.6 & 0.19\\
RG-Nr & $ 6.13\times 10^{12}$& $6.37\times 10^{-1}$& 14 & 0 & 0.01\\
RG-N4 & $ 8.58\times 10^{12}$& $5.97\times 10^{-1}$ & 23 & 0 & 0.046\\
RG-N5 & $ 1.07\times 10^{13}$& $4.99\times 10^{-1}$& 4.6 & 0 & 0.16\\
He-r & $ 3.86\times 10^{10}$& $2.3\times 10^{-2}$ & 65.6 & 522 & 0.82\\
He-4 & $ 5.43\times 10^{10}$& $8.1\times 10^{-3}$ & 28.9 & 440 & 0.74\\
He-5 & $ 6.79\times 10^{10}$& $4.11\times 10^{-3}$ & 14.7 & 394 & 0.74\\
He-Nr & $ 3.86\times 10^{10}$& $1.99\times 10^{-2}$ & 77 & 0 & 0.48\\
He-N4 & $ 5.43\times 10^{10}$& $ 7.3\times 10^{-3}$& 29 & 0 & 0.53\\
He-N5 & $ 6.79\times 10^{10}$& $3.82\times 10^{-3}$ & 15.5 & 0 & 0.56\\
He-2D-Nr & $ 3.86\times 10^{10}$& $1.73\times 10^{-2}$ & 63 & $-$ &$-$\\
He-2D-N4 & $ 5.43\times 10^{10}$& $5.3\times 10^{-3}$ & 12.7 & $-$ &$-$\\
He-2D-N5 & $ 6.79\times 10^{10}$& $2.6\times 10^{-3}$ & 6.3 & $-$ &$-$\\
\tableline
\end{tabular}
\tablenotetext{a}{The initial orbital separation, $a$}
\tablenotetext{b}{The final unbound mass, $\delta M_{\rm ub}$, described in equation~\ref{eqfub}.}
\tablenotetext{c}{The kick velocity, $\delta v_{\rm kick}$, described in equation~\ref{eqkick}.}
\tablenotetext{e}{The initial orbital speed}
\tablenotetext{e}{The stripped mass to ablated mass ratio}
\tablenotetext{\dagger}{Note: 
The cases beginning with ``MS,'' ``RG,'' and ``He'' correspond to the MS-WD, RG-WD, and He-WDc scenarios in Table~\ref{tab1}, respectively. The detailed description for the notations of simulation cases is described in Section~\ref{simulations}. Orbital motion and passive particles are not included in the
two-dimensional simulations, so for these cases the orbital velocity and stripped to ablated ratio are not calculated.}
\end{center}
\end{table}

In this section, the evolution of the companion stars during the SN~Ia 
explosion for the MS star, RG, and He-star binary companions (all the 
simulation cases are summarized in Table~\ref{tabAllruns}) is qualitatively described. 
Convergence tests of the two-dimensional and three-dimensional simulations were performed 
to determine the sensitivity of the numerical spatial resolutions.
A detailed description of the SN~Ia explosion subgrid model setup is examined as well. 

\subsection{A qualitative description of the evolution after the SN~Ia explosion \label{sec_ms_nr}}

\begin{figure}
\epsscale{0.8}
\plotone{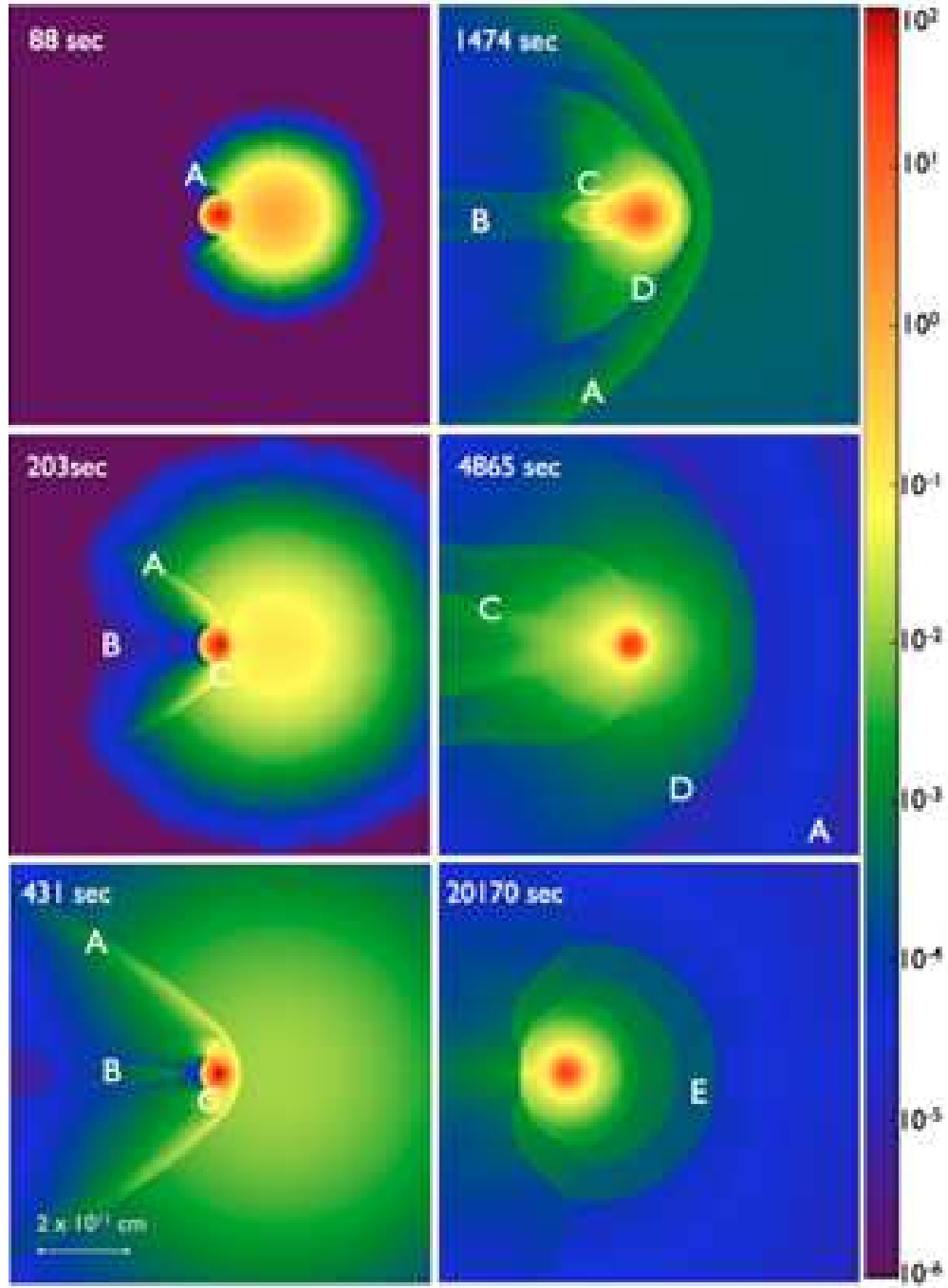}
\caption{\label{dens_ms_sn} Gas density distribution in the orbital plane 
for a three-dimensional SN~Ia simulation with a MS star binary companion 
(case MS-Nr in Table~\ref{tabAllruns}). 
The initial orbital separation is $1.51\times 10^{11}$~cm (RLOF), but orbital motion is ignored.
Each frame shows a portion of the domain spanning $20 R_*$ at the labeled simulation time.
The color scale indicates the logarithm of the gas density in g~cm$^{-3}$}.
\end{figure}

Figure~\ref{dens_ms_sn} demonstrates a typical evolution of the gas density in the orbital plane for the impact of a SN~Ia in the MS-WD scenario.
In this case (the case MS-Nr in Table~\ref{tabAllruns}), the simulation is three-dimensional, but the orbital motion is ignored.
The initial binary separation is $1.51\times 10^{11}$~cm, $\sim 3 R_*$, and 6/8 levels of refinement are used.
At $88$~sec, the SN ejecta reaches the MS star companion and forms a bow shock at the leading surface of the MS star (label A in Figure~\ref{dens_ms_sn}).
The bow shock extends further, creating an opening angle with $\sim 40^\circ$ with respect to the $x-$axis.
As the shock propagates further, the ejecta sweep around the MS star. 
At about 200~seconds, the ejecta fully surround the MS star and self-interact at the back side of the MS star (label B Figure~\ref{dens_ms_sn} at $203$~sec). 
Subsequently, a cone-shape tail shock forms (label B at $431$~sec).  
The shock waves also penetrate the MS star (label C in Figure~\ref{dens_ms_sn}), reaching the center at about $431$~seconds.
When the shock passes through the MS star, a reverse shock begins to reflect and refill the central SN region (label D in Figure~\ref{dens_ms_sn}).
The reverse shock mainly contains companion material and forms a solid angle spanning about $40^\circ-50^\circ$ with respect to the $x-$axis.
After the impact, the MS star pulsates, creating shocks around the surface of the MS star (label E in Figure~\ref{dens_ms_sn}). 
By the end of the simulation, $0.18 M_\odot$ of mass has been lost (gravitationally unbound) due to the impact of the SN~Ia. The MS star also receives linear momentum and hence a kick of $112$~km/sec from the ejecta.

Qualitatively, the simulation resembles the results in \cite{2000ApJS..128..615M} and \cite{2008A&A...489..943P}.
The case rp3\_20a in \cite{2008A&A...489..943P} resembles ours, except that the orbital separation is smaller ($a=2.68\times 10^{11}$~cm in \cite{2008A&A...489..943P}).
The clearer shock structures reveal typical behavior when comparing grid-based hydro and SPH simulations.
The most significant difference is the absence of a reverse shock in the SPH simulations.
Although \cite{2008A&A...489..943P} demonstrated that their SPH code could produce the same amount of unbound mass as a grid-based hydro code, we believe that the high-speed reverse shock actually plays a more important role for the dynamics of the remnant star as it could take away some momentum from the SN ejecta.

\subsection{Effects due to the subgrid SN~Ia setup}

\begin{figure}
\plotone{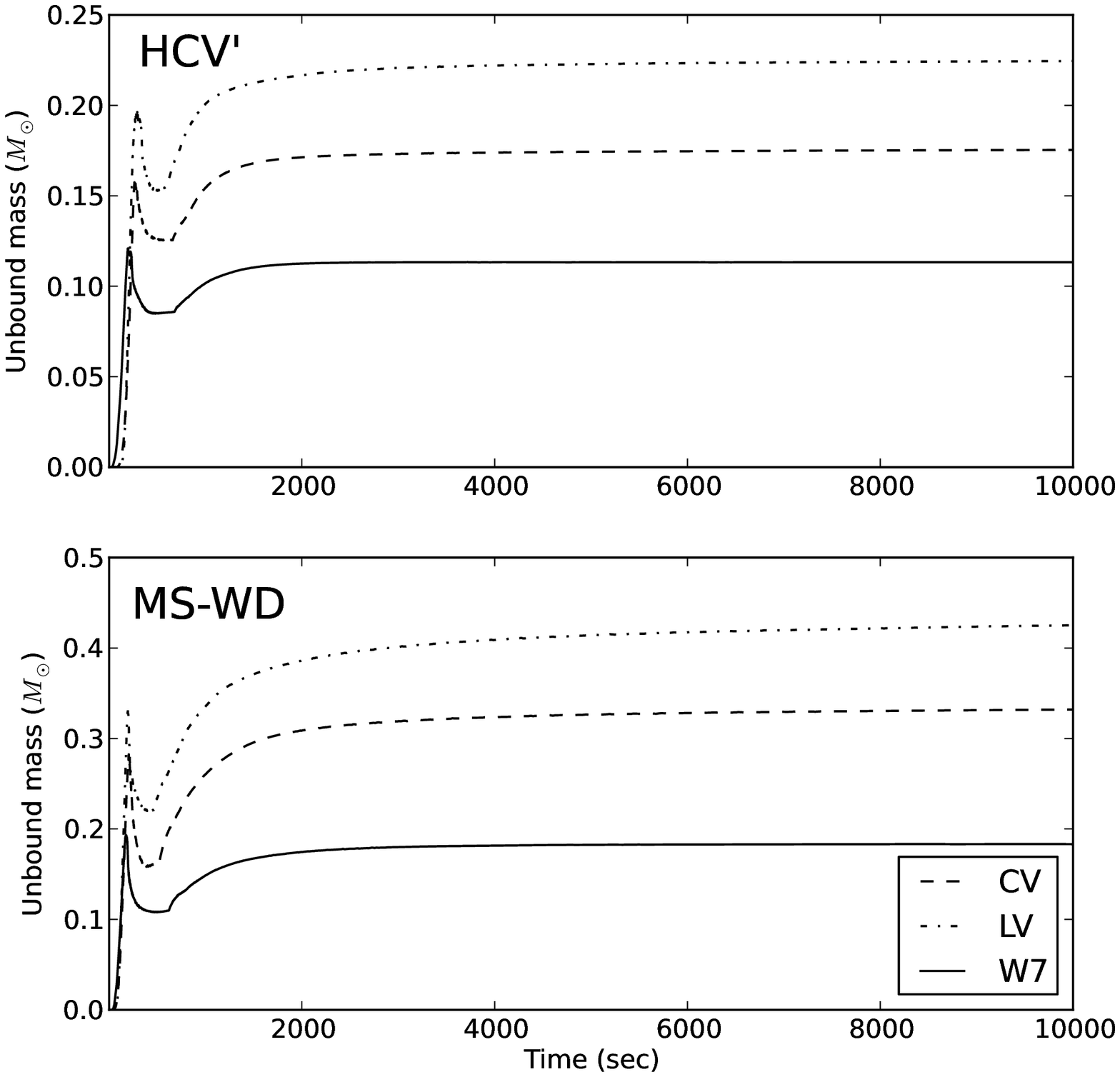}
\caption{\label{snmodel} Unbound mass vs. simulation time for different SN explosion setups (UPPER: HCV'; LOWER: MS-WD). 
CV denotes the case with constant density and radial velocity distribution; LV denotes the case in which the velocity is linearly increased in radius, but the density distribution is uniform; 
W7 denotes the fitted model described in section~\ref{sec_snmodel}. The simulations are two-dimensional, and the binary separation is $2.03 \times 10^{11}$~cm ($= 3 R_*$) for the HCV' scenario and $1.51 \times 10^{11}$~cm ($\sim 2.74 R_*$) for the MS-WD scenario.}
\end{figure}

In SNe~Ia, hydrogen has not been detected. However, hydrogen-rich material from the accretion disk or from mass stripping during the SN impact is possible and could be detectable when the SN ejecta become transparent. 
Thus, one of the most significant quantities we wish to determine is the amount of mass lost from the hydrogen-rich companion star after the SN~Ia explosion. 
The unbound mass could be found by calculating the difference between the initial mass of the 
companion star and the integrated total bound mass from the companion at each timestep. 
The total bound mass is the sum of the companion cell masses in all zones for which the total energy is negative.

Because of the limitation of spatial resolution, the explosion cannot be perfectly modeled.
It is found that the amount of unbound mass is sensitive to the sub-grid model representing the SN~Ia.
Therefore, a realistic description of the SN~Ia explosion is required in addition to including the correct amount of energy.
 
\cite{1998ApJ...497..807D, 2000ApJ...541..418D} examined the interaction of SN~Ia ejecta with the surrounding interstellar medium, suggesting a density profile characterized by an exponential for setting up the SN~Ia explosion.
However, our simulations are restricted to a much smaller scale ($10^{10}$~cm to $10^{13}$~cm rather 
than $10^{15}$~cm); the binary companion will break the self-similarity of the SN explosion. 
Thus, the exponential decay may not be appropriate here.
We experimented with several different initial distributions of SN~Ia in order to study their effect on the post-impact companion star. 

The most simple and naive model is to set up a uniform density and constant radial velocity distribution in the SN region (denoted by CV SN). 
However, in reality, the beginning of the explosion looks more like a Sedov explosion. 
Thus, a second model is set up with the radial velocity distribution linearly increasing with radius; 
the gas density is assumed to be uniformly distributed (denoted by LV SN).
We calculated the amount of unbound mass after the SN~Ia explosion for these models and found that the different SN~Ia models will result in a difference of unbound mass of about $\sim 0.05-0.3 M_{\odot}$ (see Figure~\ref{snmodel}).

Thus, a more realistic sub-grid model is necessary to model the SN~Ia explosion.
The early time evolution and distribution of the W7 model is well described in \cite{1984ApJ...286..644N}. 
We approximate the density profile in Figure~4b of \cite{1984ApJ...286..644N} by following a power-law distribution with a fixed slope in the $M-r$ relation when the explosion expands to the size of our sub-grid model, $R_c$.
For a given total mass ($M_{\rm WD}$) and an explosion size ($R_{\rm c}$), the density profile can be calculated by
\begin{equation}
\rho(r) = \frac{M_0}{r^3+\zeta R^3_{\rm c}}
\end{equation}
where $M_0=2.4998 \times 10^{32}$~g is a constant only related to the slope of the $M-r$ relation in \cite{1984ApJ...286..644N}. 
The other constant,
\begin{equation}
\zeta= \left( e^{\frac{3 M_{\rm WD}}{4 \pi M_0}}-1 \right) ^{-1}
\end{equation}
depends on $M_0$ and $M_{\rm WD}$.
The velocity distribution also is assumed to increase linearly in radius, $v(r)= f v_{\rm sn} (r/R_{\rm c})$,  which is a reasonable assumption and comparable with Figure~4c of \cite{1984ApJ...286..644N}.  
If we know the density distribution and the total kinetic energy, the coefficient of the velocity profile, $f$, can be calculated by integrating the kinetic energy density.
Based on the temperature profile in Figure~4a of \cite{1984ApJ...286..644N}, 
we assume the temperature is uniformly distributed. 
We use the Helmholtz EOS solver to iterate the temperature until the total explosion energy is equal to $1.233 \times 10^{51}$~erg (denote by W7 SN).

Figure~\ref{snmodel} illustrates two-dimensional simulations of the unbound mass versus simulation time with different SN~Ia models and different companion models (HCV' and MS-WD).
For the HCV' companion model, the final unbound mass is $0.113 M_{\odot}$ for the W7 SN model, $0.175 M_{\odot}$ for the CV SN model, and $0.225M_{\odot}$ for the LV SN model. 
\cite{2000ApJS..128..615M} obtained $0.15M_{\odot}$ by using their HCV model; \cite{2008A&A...489..943P} using the same model but with an SPH simulation, obtained $0.134 M_{\odot}$.
Their final unbound mass is between our CV SN model and W7 SN model. 

For the other companion model, the MS-WD scenario, we obtained $0.179M_{\odot}$ for the W7 SN model, $0.330 M_{\odot}$ for the CV SN model, and $0.427M_{\odot}$ for the LV SN model. 
\cite{2008A&A...489..943P} used the same companion from \cite{2004ApJ...601.1058I}; due to the difference of constructing this model, \cite{2008A&A...489..943P} has a slightly larger radius and different material compositions. 
From their fitted power-law relation, the final unbound mass at our initial binary separation ($a=2.74 R_*$) is $0.19 M_{\odot}$, which is consistent with our results.
We therefore use the W7~SN model for all other cases in this paper.

\subsection{Convergence Test}

\begin{figure}
\plotone{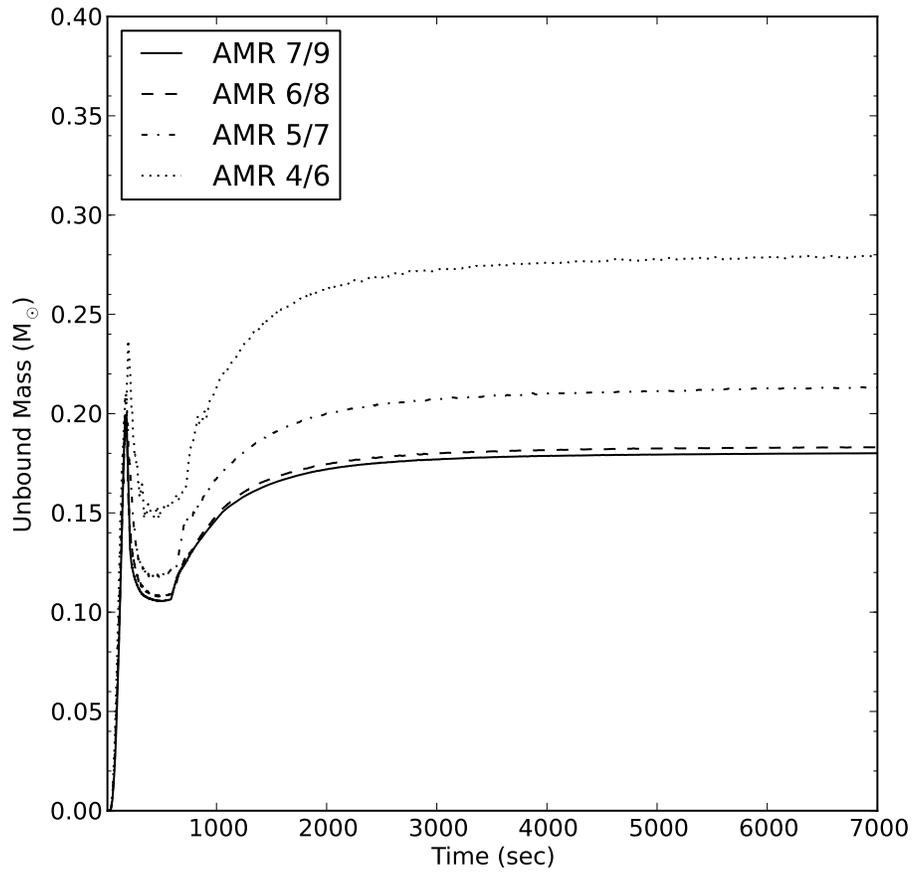}
\caption{\label{convergence2d} Unbound mass from the companion star vs. simulation time using different AMR refinement levels for the MS-WD scenario. The initial binary separation is $1.51 \times 10^{11}$ cm and the calculation is two-dimensional. }
\end{figure}
%

\begin{figure}
\plotone{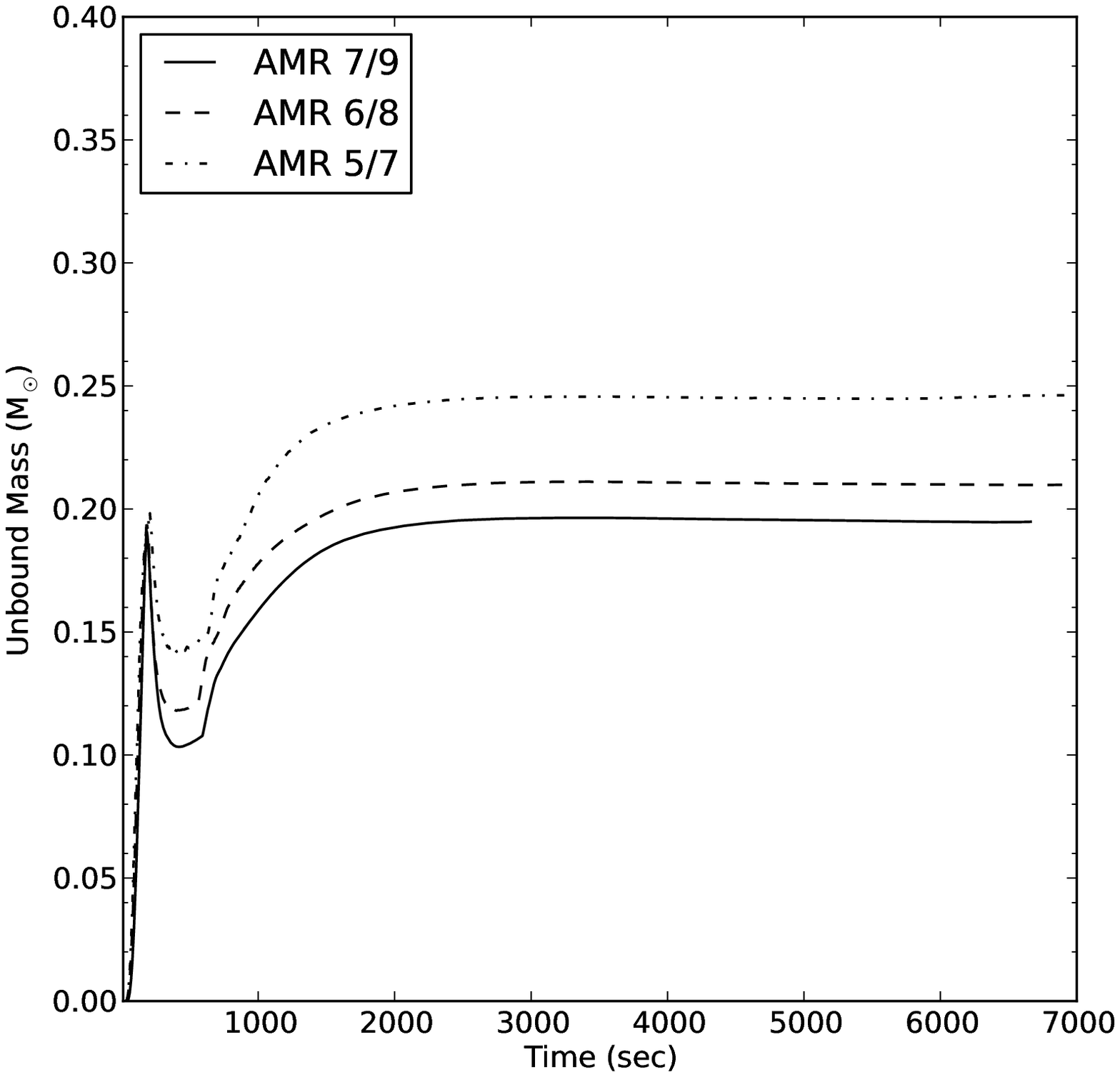}
\caption{\label{convergence3d} Similar to Figure\ref{convergence2d}, but with three-dimensional simulations, including orbital motion and spin. }
\end{figure}

In order to check the sensitivity of unbound mass to spatial resolution, we performed convergence tests on two-dimensional and three-dimensional simulations using the MS-WD companion model. 
We performed several simulations with the same initial binary separation ($a=1.51\times 10^{10}$~cm), and the same W7 SN model for the SN~Ia explosion, but with several different AMR levels, computing the amount of unbound mass as a function of time for each simulation.
In Figure~\ref{convergence2d}, we see that the unbound mass decreases when the zone spacing decreases; the decrement changes very little from $6/8$ to $7/9$ levels. 
The difference between $6/8$ (corresponding to $1024^3$ in 3D and $1024\times 2048$ in 2D) and $7/9$ (corresponding to $2048^3$ in 3D and $2048\times 4096$ in 2D) levels is within $3\%$.
The final unbound mass of this two-dimensional run for the MS-WD scenario is $M_{\rm ub}=0.175 M_\odot$. 

For the convergence test of three-dimensional simulations, we use the same companion model (MS-WD), but the spin and orbital motion are included.
In this stage, we are only interested in the sensitivity to numerical resolution during the SN~Ia explosion. 
Thus, we ignored the effects of accretion of Roche-filling material, i.e. the WD exploded immediately when it was added to the simulation. 
The result is shown in Figure~\ref{convergence3d}.
The convergence behavior is the same as for two-dimensional simulations. 
However, three-dimensional runs require a higher AMR level for the same spatial resolution because the cylindrical mesh has only half of the zone spacing, due to symmetry, of the Cartesian mesh for the same number of AMR levels. 
The final unbound mass $M_{\rm ub}=0.195 M_\odot$ for $7/9$ levels. 
The difference between $6/8$ and $7/9$ is about $8\%$, which is larger than the difference in the two-dimensional case.
However, a full set of runs at $7/9$ levels proved to be too expensive for us to undertake. 
Thus, we chose $6/8$ levels for all other production runs.

\subsection{Evolution of the Companion Star after the SN~Ia Explosion \label{simulations}}

We have performed several simulations with the MS-WD, RG-WD, and He-WDc companions (see Table~\ref{tab1}).
For each scenario, we ran cases with and without the orbital motion to distinguish the effects of asymmetry. 
We also compared simulations between three dimensions and two dimensions. 
In addition, a parameter survey of changing initial orbital separations was also carried out. 

Table~\ref{tabAllruns} summarizes all our numerical simulations.
The cases beginning with ``MS,'' ``RG,'' and ``He'' correspond to the MS-WD, RG-WD, and He-WDc scenarios in Table~\ref{tab1}, respectively.
The letter ``r'' represents the cases with initial orbital separation equal to the separation for RLOF (see equation~\ref{RLOF}); 
``4'' and ``5'' are cases with initial binary separations equal to $4R_*$ and $5R_*$. 
The corresponding orbital speeds are calculated for initially circular orbits using Kepler's third law.
The letter ``N'' represents cases that ignored the orbital motion and spin.
The cases with ``2D'' are two-dimensional simulations; others are simulated in three dimensions.

\subsubsection{Main-sequence binary companion}

\begin{figure}
\plotone{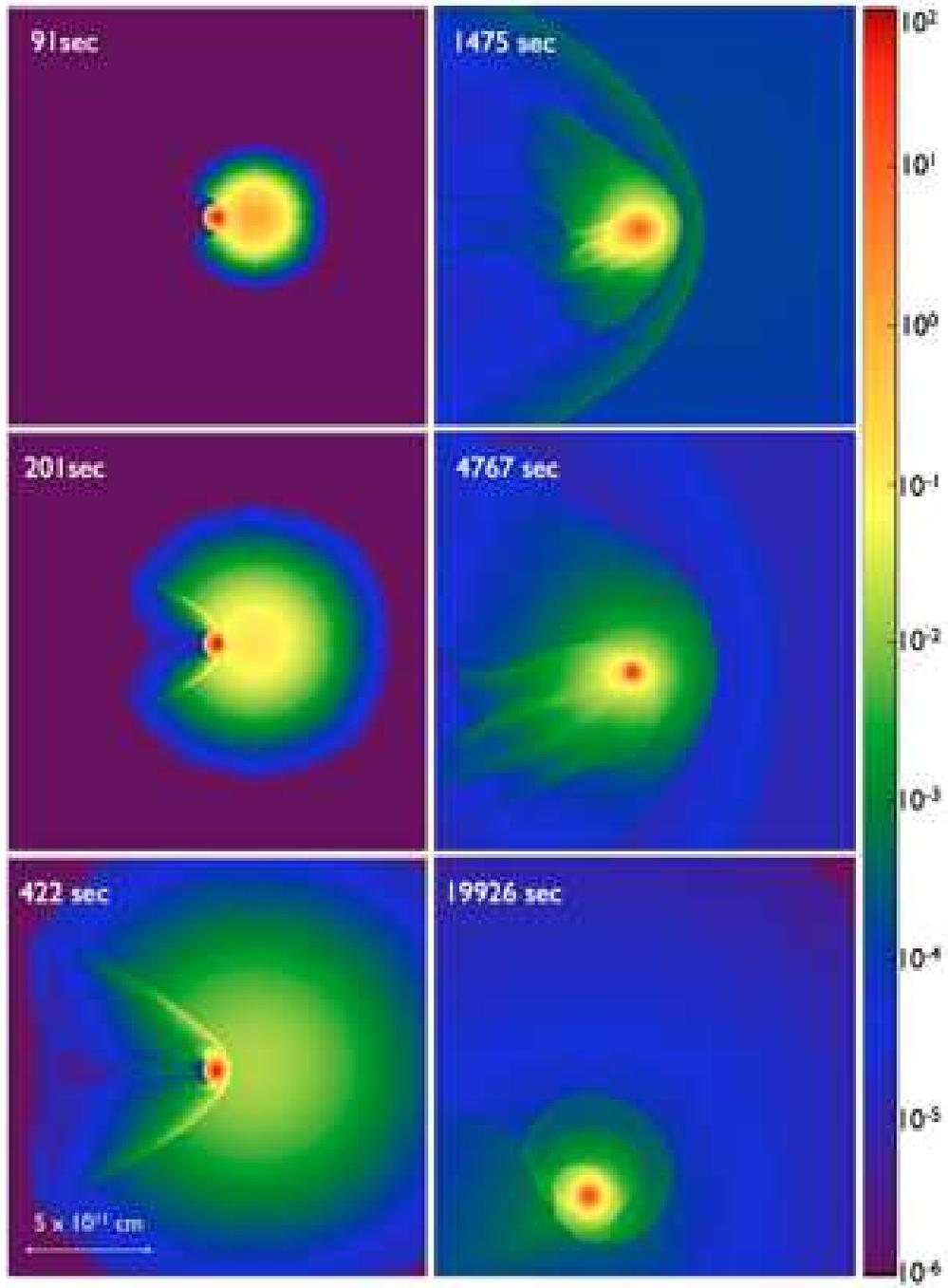}
\caption{\label{dens_ms_snbd} Similar to Figure~\ref{dens_ms_sn}, but for case MS-r; each frame reveals a portion of the domain spanning $30 R_*$.}
\end{figure}

In Section~\ref{sec_ms_nr}, we described the case MS-Nr (see Table~\ref{tabAllruns}) for a three-dimensional simulation with the MS-WD companion but without orbital motion or spin.
Figure~\ref{dens_ms_snbd} shows a similar simulation for the case MS-r, which includes the orbital motion and spin.
The simulation resembles the case MS-Nr during the first few hundred seconds, except that the tail shock is shifted and curved because the ejecta speed is much higher than the orbital 
speed.  However, after a thousand seconds, the asymmetry becomes more obvious; most of the same features, however, can still be seen.
This asymmetry will become more important for the evolution of the SNR.
Including orbital motion and spin leads to a greater unbound mass by about $16\%$, 
but the kick velocity remains the same.
The additional unbound mass may be due to the spin; the rotational energy makes fluid elements easier to unbind.   
However, the momentum lost in this additional mass may not change the center-of-mass velocity.
A more detailed description and explanation is given in Section~\ref{sec_parm}.

\subsubsection{Red-giant binary companion}

\begin{figure}
\plotone{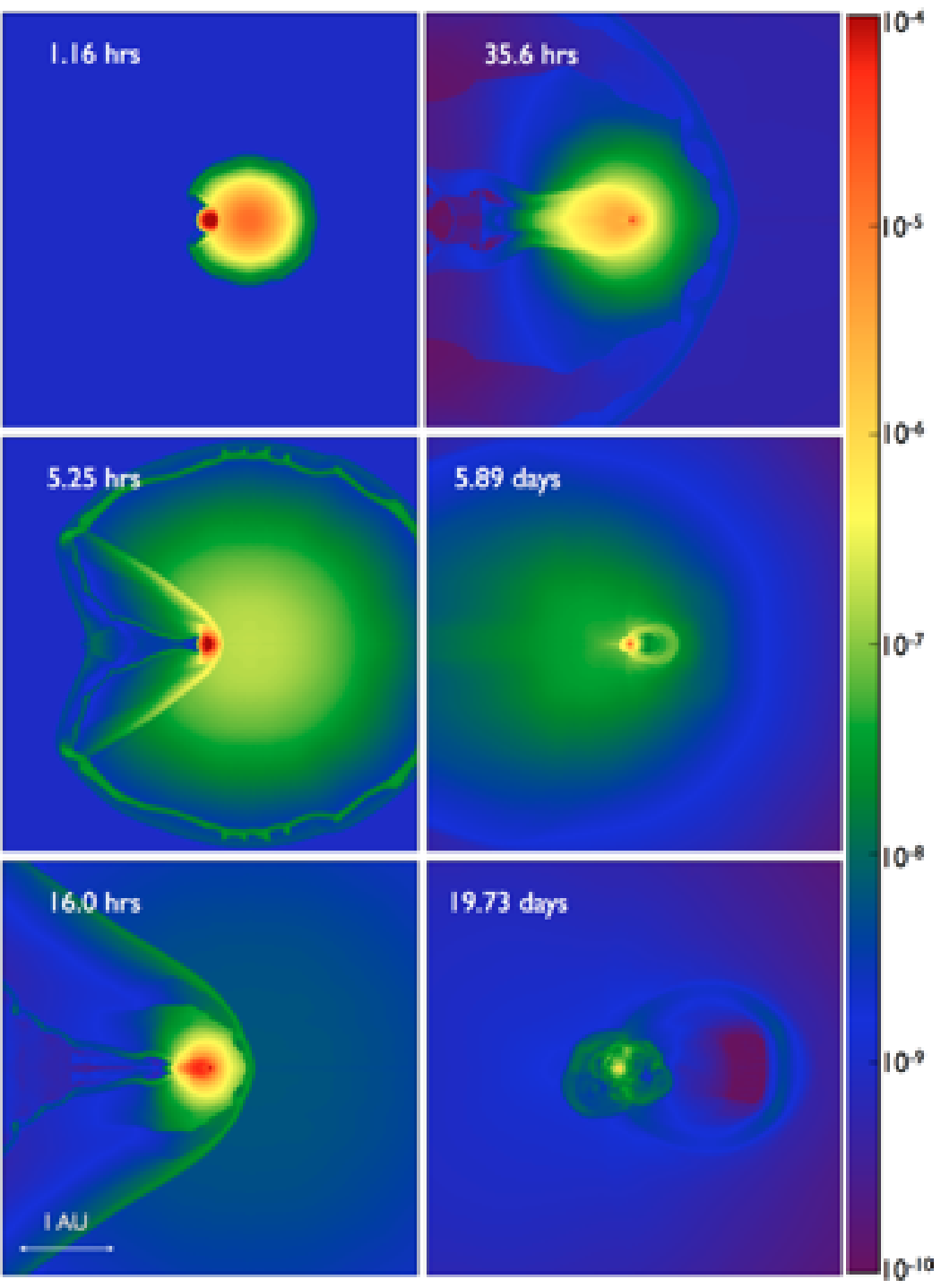}
\caption{\label{dens_rg_sn} Similar to Figure~\ref{dens_ms_snbd}, but for case RG-Nr.}
\end{figure}
%
\begin{figure}
\plotone{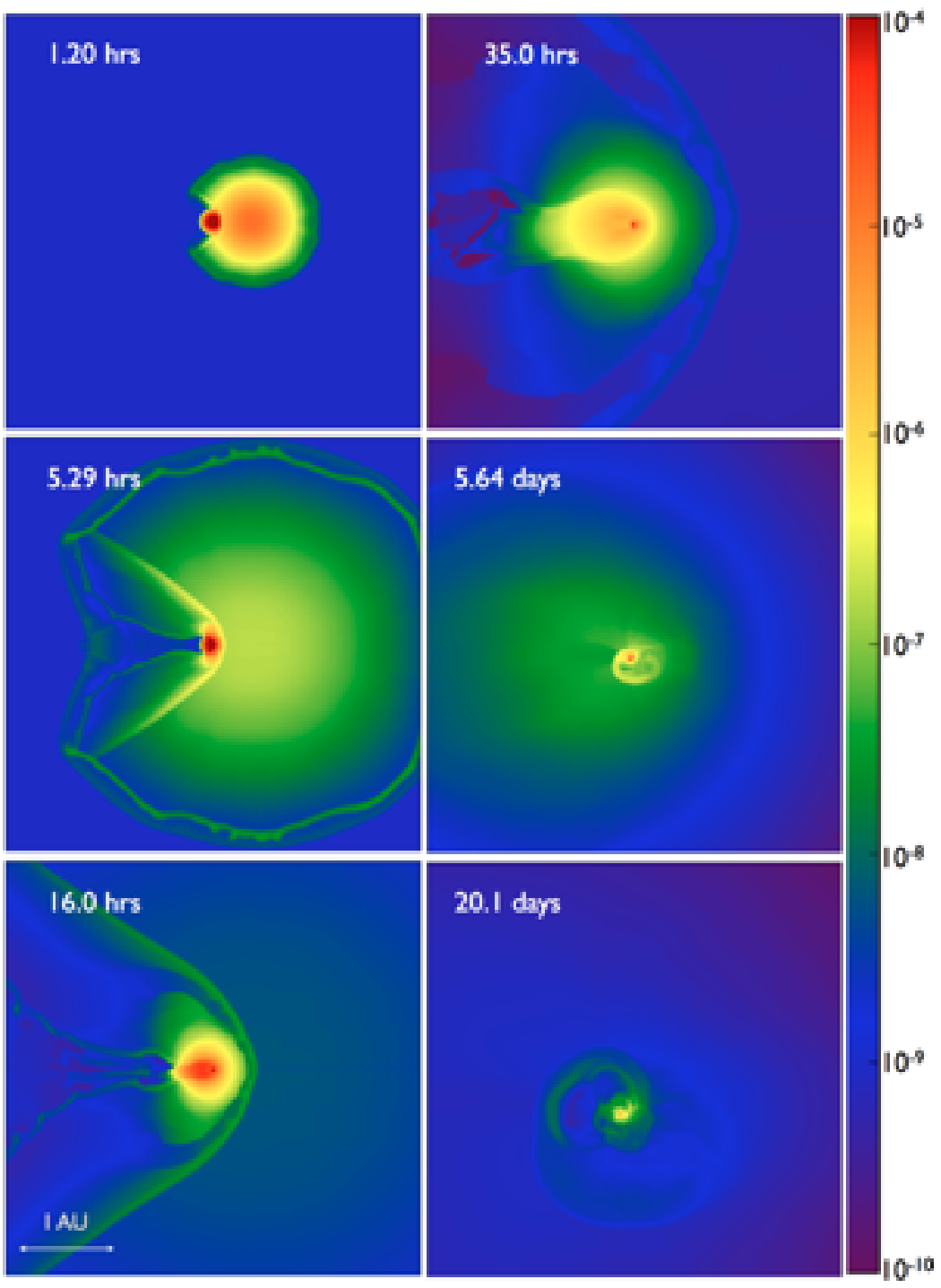}
\caption{\label{dens_rg_snbd} Similar to Figure~\ref{dens_ms_snbd}, but for case RG-r.}
\end{figure}

The evolution of gas density for the RG-WD companion model is shown in Figure~\ref{dens_rg_sn} (RG-Nr, without orbital motion) and Figure~\ref{dens_rg_snbd} (RG-r, with orbital motion).
The asymmetric effect is insignificant for the RG companion, because the orbital speed is much lower than the ejecta speed.

Unlike the MS star, almost all the envelope ($>95\%$) of the RG is removed during the initial impact.
Although the ejecta speed is smaller than for the MS-WD companion when it reaches the RG, the 
greater extent and lower binding energy of the RG makes it difficult to survive the SN~Ia explosion.
As a result, most of the momentum from the SN ejecta is transferred to the unbound mass,
creating only a tiny kick on the RG core (less than $30$~km/sec).
The bow shock has an opening angle of about $40^\circ-45^\circ$ with respect to the $x-$axis. 
For the RG-r case, a spiral pattern appears due to the combination of fallback and spin about 5~days after the SN~Ia explosion.
The reverse shock also is weaker than in the MS-WD scenario.
We can only report a lower limit of the final unbound mass: the core of the RG is represented by a particle cloud and the mass of the particle cloud is not lost in our simulation.

\subsubsection{Helium star binary companion}

\begin{figure}
\plotone{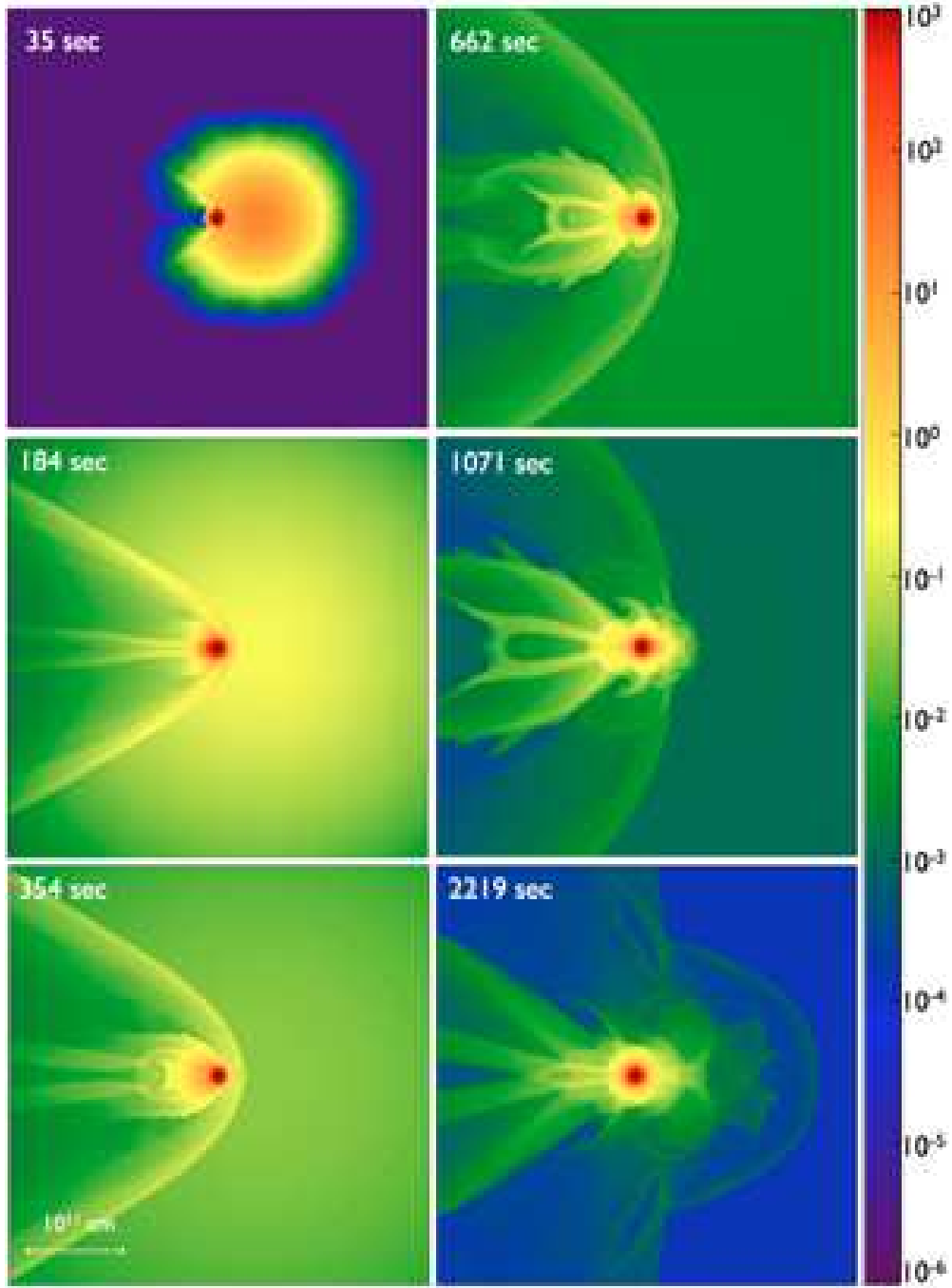}
\caption{\label{dens_he_sn}Similar to Figure~\ref{dens_ms_snbd}, but for case He-Nr.}
\end{figure}

\begin{figure}
\plotone{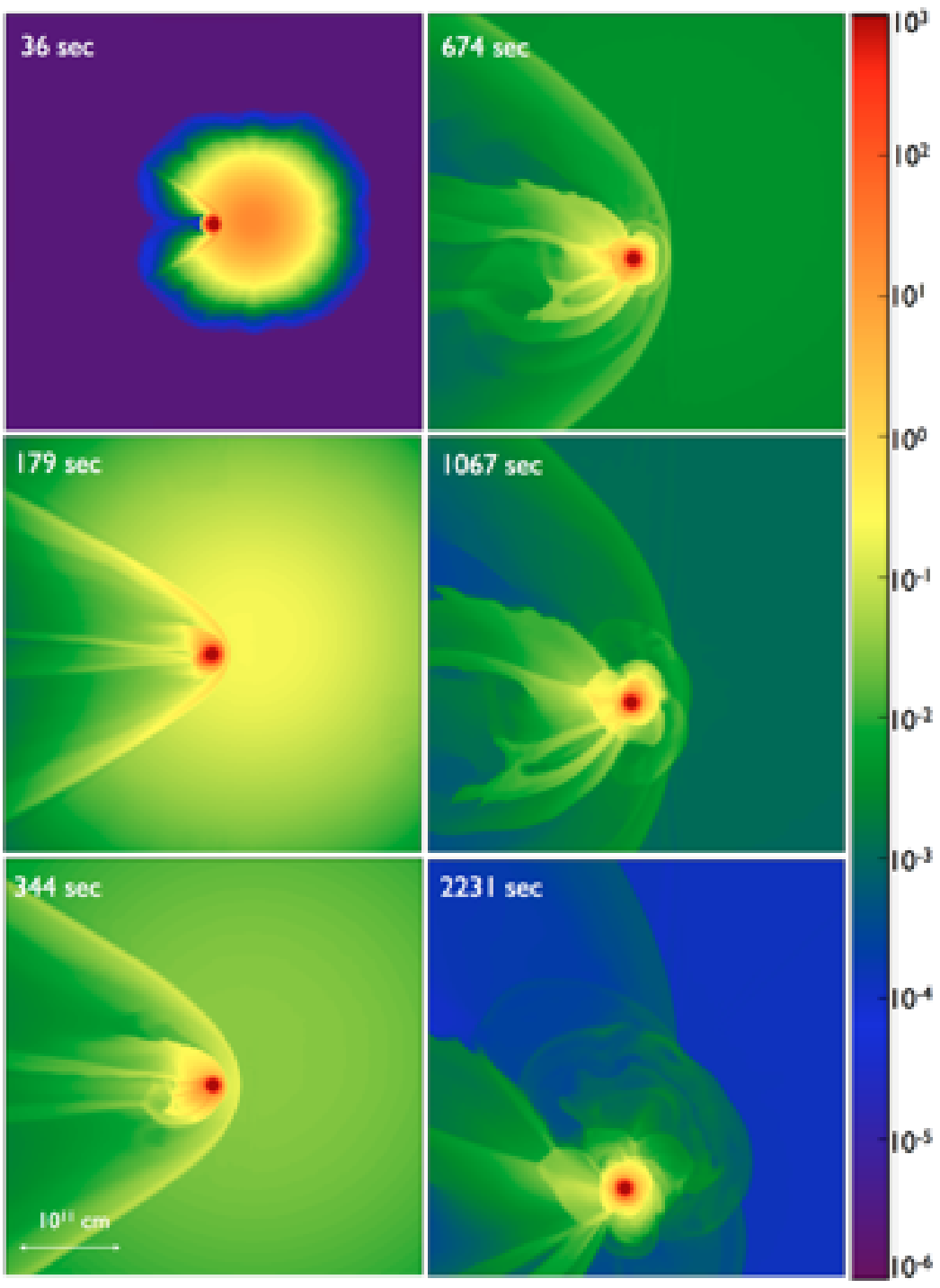}
\caption{\label{dens_he_snbd}Similar to Figure~\ref{dens_ms_snbd}, but for case He-r.}
\end{figure}

The He-WDc companion has the most dramatic impact and is the most compact companion in comparison to the MS-WD and RG-WD cases.
Figure~\ref{dens_he_sn} and Figure~\ref{dens_he_snbd} show the evolution of gas density in the orbital plane for the He-Nr and He-r cases.
The opening angle of the bow shock is about $30^\circ-35^\circ$.
Because the orbital speed is higher than in the MS-WD and RG-WD scenarios, asymmetric effects due to the orbital motion appear earlier (within the first few hundreds of seconds).

\begin{figure}
\plotone{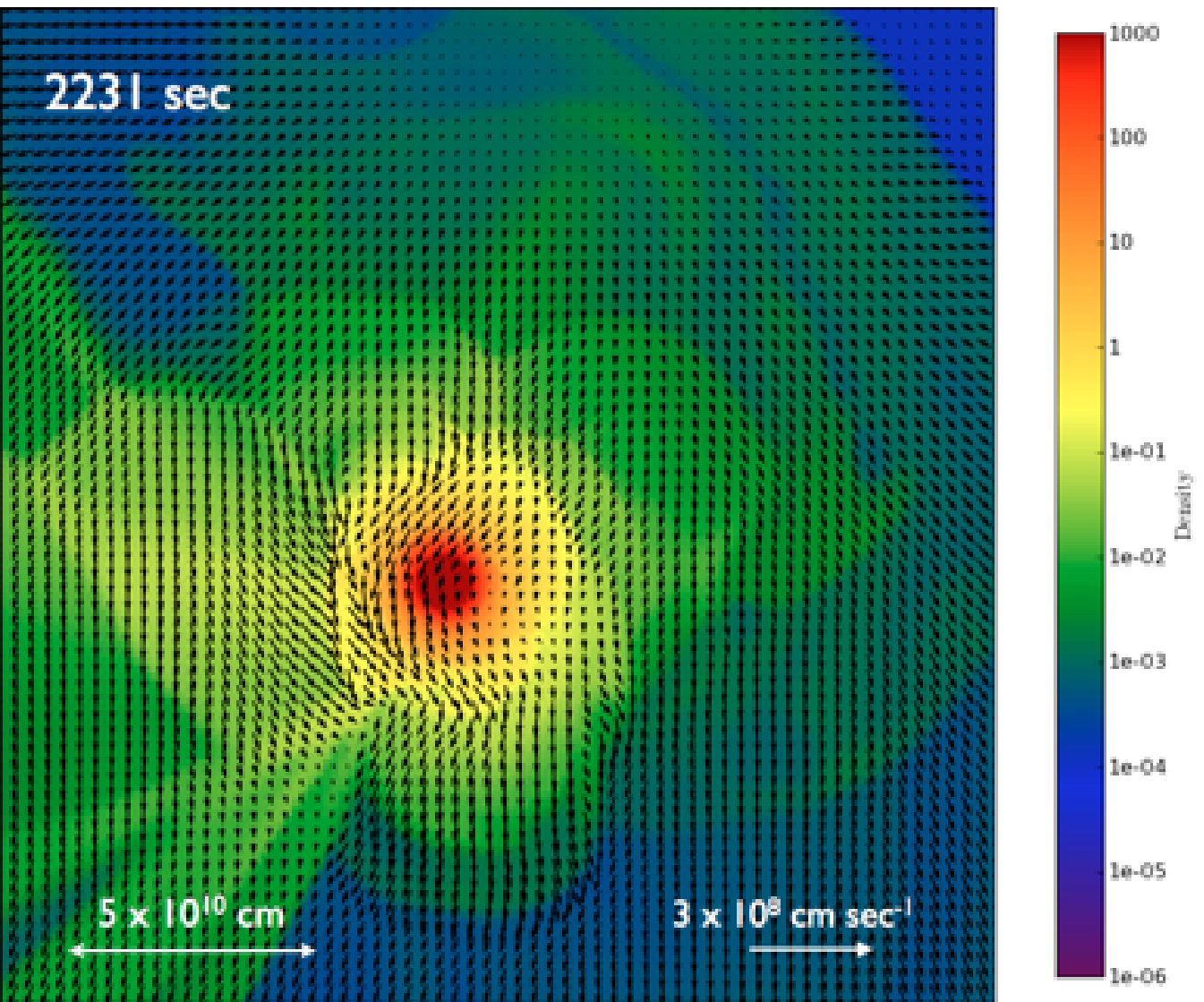}
\caption{\label{vfield} Velocity vector field around the helium star companion at 2231~sec after the SN~Ia explosion in the simulation He-r. The velocity vectors are plotted every 10 zones. The white arrows reveal the scale of the domain and the scale of the velocity vectors.
The color displays the logarithm of the gas density in g~cm$^{-3}$ in the orbital plane.}
\end{figure}

After the initial impact, the compressed helium star oscillates. 
The oscillation frequency is higher and the flow is more turbulent in comparison to the MS companion.
After several hundreds of seconds, the strong velocity shear behind the helium star leads to Kelvin-Helmholtz instability. 
Later, the oscillation of the helium star generates shocks which interact with Rayleigh-Taylor instabilities owing to the gravity from the helium star.
Therefore, at the end of the simulation, a turbulent remnant environment is obtained.
Figure~\ref{vfield} shows the velocity vector field in a gas density slice at $2231$~sec in the orbital plane for the case He-r.
There is a global kick velocity in the lower-left direction, caused by the orbital speed 
and a random turbulent velocity.


\section{Discussion  \label{sec_disc}}

\subsection{Parameter Survey \label{sec_parm}}

\begin{table}
\begin{center}
\caption{Power-law indices for the final unbound mass and kick velocity. \label{tab3}}
\begin{tabular}{lcccc}
\\
\tableline
Model          & $m_{\rm ub}$ \tablenotemark{a}   & $C_{\rm ub}$ \tablenotemark{a}  &  $m_{\rm kick}$\tablenotemark{b}  & $C_{\rm kick}$ \tablenotemark{b} \\
\tableline
MS            & $-3.61$   & $0.894$ & $-1.76$& $2.83$\\
MS-N        & $-3.46$   & $0.757$ & $-1.69$& $2.78$\\
MS-2D-N  & $-3.66$   & $0.859$ & $-1.88$& $2.89$ \\
RG           & $-0.391$ & $-0.0103$ & $-$ & $-$\\
RG-N       & $-0.416$ & $0.00269$ & $-$ & $-$\\
He           & $-3.04$   & $-0.260$ & $-2.62$& $3.02$\\
He-N       & $-2.92$ & $-0.381$ & $-2.83$& $3.17$\\
He-2D-N & $-3.36$ & $-0.245$ & $-4.6$& 3.9\\
\tableline
\end{tabular}
\tablenotetext{a}{The power-law index, $m_{\rm ub}$, and constant, $C_{\rm ub}$, described in equation~\ref{eqfub}.}
\tablenotetext{b}{The power-law index, $m_{\rm kick}$, and constant, $C_{\rm kick}$, described in equation~\ref{eqkick}.}
\end{center}
\end{table}

\begin{figure}
\plotone{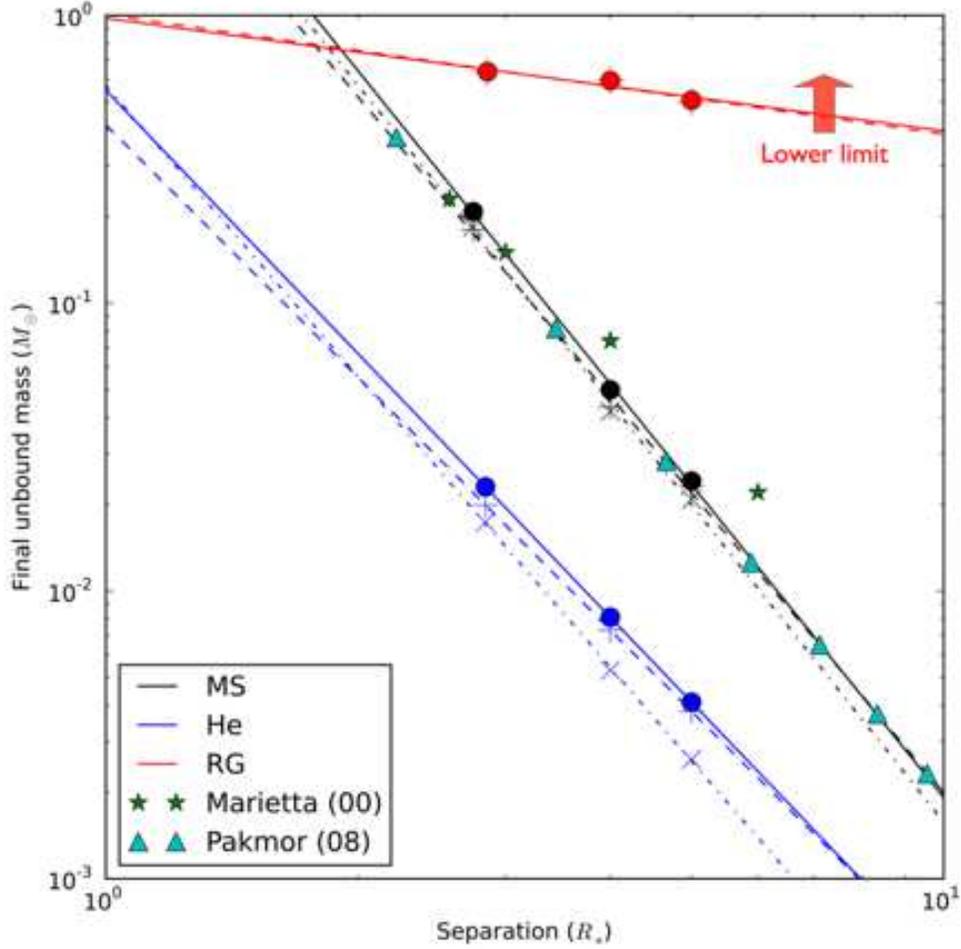}
\caption{\label{fubSep} Final unbound stellar mass vs. binary separation for different companion models. 
Each color represents data for a type of companion model: black represents simulations for the MS-WD scenario; 
blue, the He-WDc scenario; 
and red, the RG-WD scenario. 
The separation is expressed in units of the companion radius.
Lines show fitted power-law relations based on the numerical simulation results 
(the red line shows the lower limit of the RG-WD scenario).
The "$+$" symbols and dashed lines indicate 3D data without orbital motion; 
"o" symbols and solid lines present 3D data with orbital motion; 
and "$x$" symbols and dashed dot lines demonstrate 2D data without orbital motion.
Star symbols show the data from \cite{2000ApJS..128..615M} for the HCV model. 
Triangles indicate the power-law relation for a MS binary companion from \cite{2008A&A...489..943P}. }
\end{figure}
\begin{figure}
\plotone{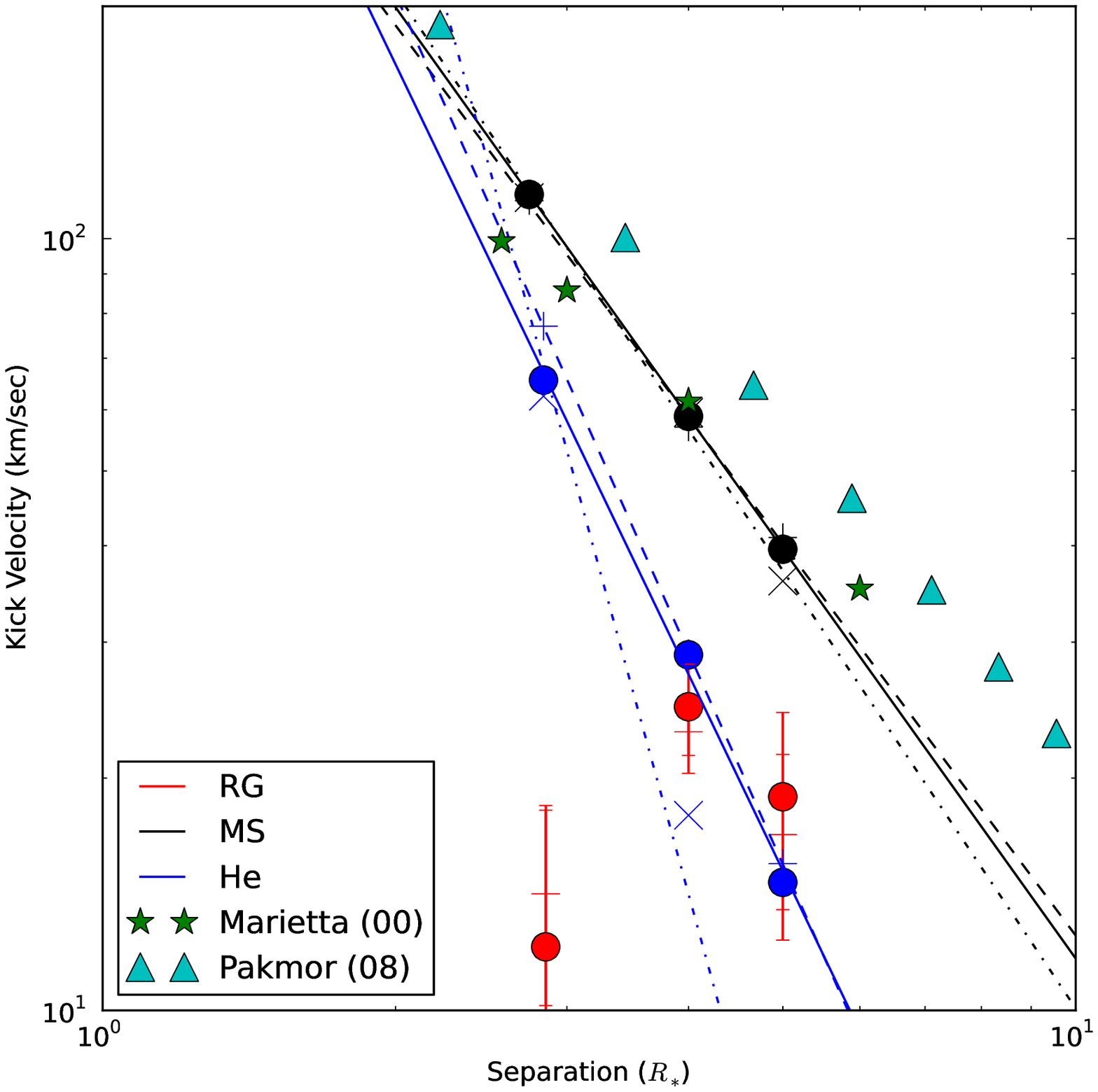}
\caption{\label{vkickSep} Similar to Figure~\ref{fubSep}, but for kick velocity vs. binary separation in different companion models. }
\end{figure}

Although most SDS channels assume the binary system is in RLOF, the actual evolutionary stage and explosion conditions may differ from the model we used in our simulations.   
We therefore have performed a parameter survey to explore the dependence of the numerical results on varying binary separation. 
Significant physical quantities for each run are summarized in Table~\ref{tabAllruns}.

Figure~\ref{fubSep} and Figure~\ref{vkickSep} show the final unbound mass and companion kick velocity as functions of the binary separation in units of companion radius.
Note that the unbound masses for the RG-WD scenario are only lower limits, because the unbound mass is still increasing at the end of these simulations.
In addition, the core of the RG is represented by a rigid particle cloud for which the mass is assumed to be unchanged during the simulation.
Therefore, we effectively ignored interactions such as mass stripping and ablation on the core of the RG. 
The companion kick velocity can be derived from the difference between the companion center-of-mass velocity at the end of a simulation and the velocity at the time when the ejecta reached the companion star in that simulation. 

In general, the final unbound mass and kick velocity decrease when increasing the binary separation.
We find that the final unbound mass, $\delta M_{ub}$, can be described by the relation
\begin{equation}
\delta M_{\rm ub} = C_{\rm ub} a^{m_{\rm ub}} M_\odot, \label{eqfub}
\end{equation}
where $a$ is the orbital separation, $m_{\rm ub}$ is the power-law index, 
and the constant $C_{\rm ub}$ depends on the companion model (see Table~\ref{tab3}).
All the companion models we used could be described with this power-law relation; the power-law indices, however, are different for different companion models.
 
Although the initial binary separation is the smallest for the He-WD scenario, suggesting the impact of the SN ejecta for this scenario should be the largest, the He star is more compact and has a higher binding energy than the MS star or RG.
Therefore the amount of unbound mass is smallest in the He-WD scenario.
In the RG-WD scenario, the gravitational force acting on the core region is only approximate since 
the softening length in the particle cloud representing the core is comparable to the size of the remaining envelope.
However, most of the envelope is lost with at least $75\%-96 \%$ ejected depending 
on the binary separation.

For the MS-WD scenario, the difference between two-dimensional and three-dimensional runs varies from $3\%$, for the smallest binary separations, to $10\%$, for the largest binary separations.
The error in the unbound mass is usually higher for the larger binary separations because the unbound mass decreases by a power-law relation with a high power-law index. 
The results between 2 and 3 dimensions in the He-WD scenario are even larger 
than in the MS-WD scenario.  
The uncertainty arises not only from the low 
percentage of unbound mass for the He star, but also from the high amount of turbulence seen in these models.
\cite{2010ApJ...715...78P} demonstrated that the effect of the turbulence precludes convergence of the unbound mass no matter the spatial resolution  and leads to a $14\%$ error for 
the He-WD scenario in the two-dimensional simulations. 

\begin{figure}
\epsscale{1.0}
\plotone{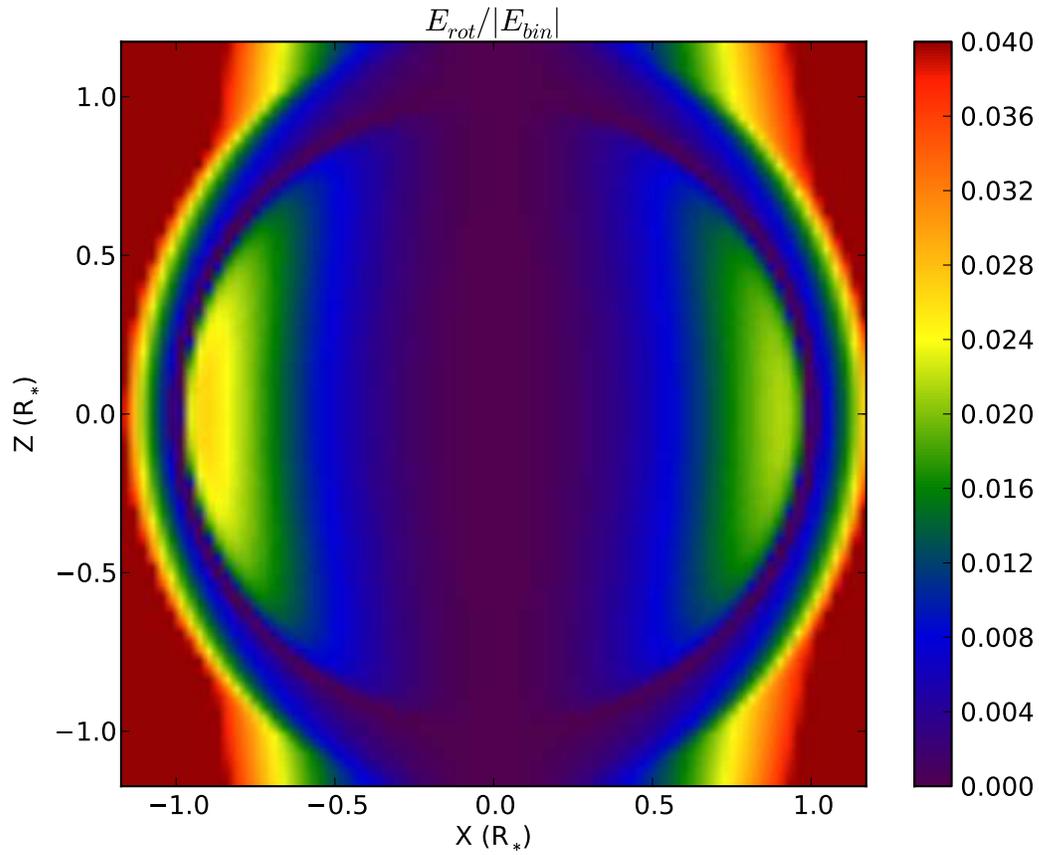}
\caption{\label{rotEner} The distribution of the ratio of the rotational energy density and binding energy density in the $x-z$ plane for the MS-r run. The region is rescaled to the units of the companion star ($R_*=5.51\times 10^{10}$~cm).}
\end{figure}
\begin{figure}
\epsscale{0.55}
\plotone{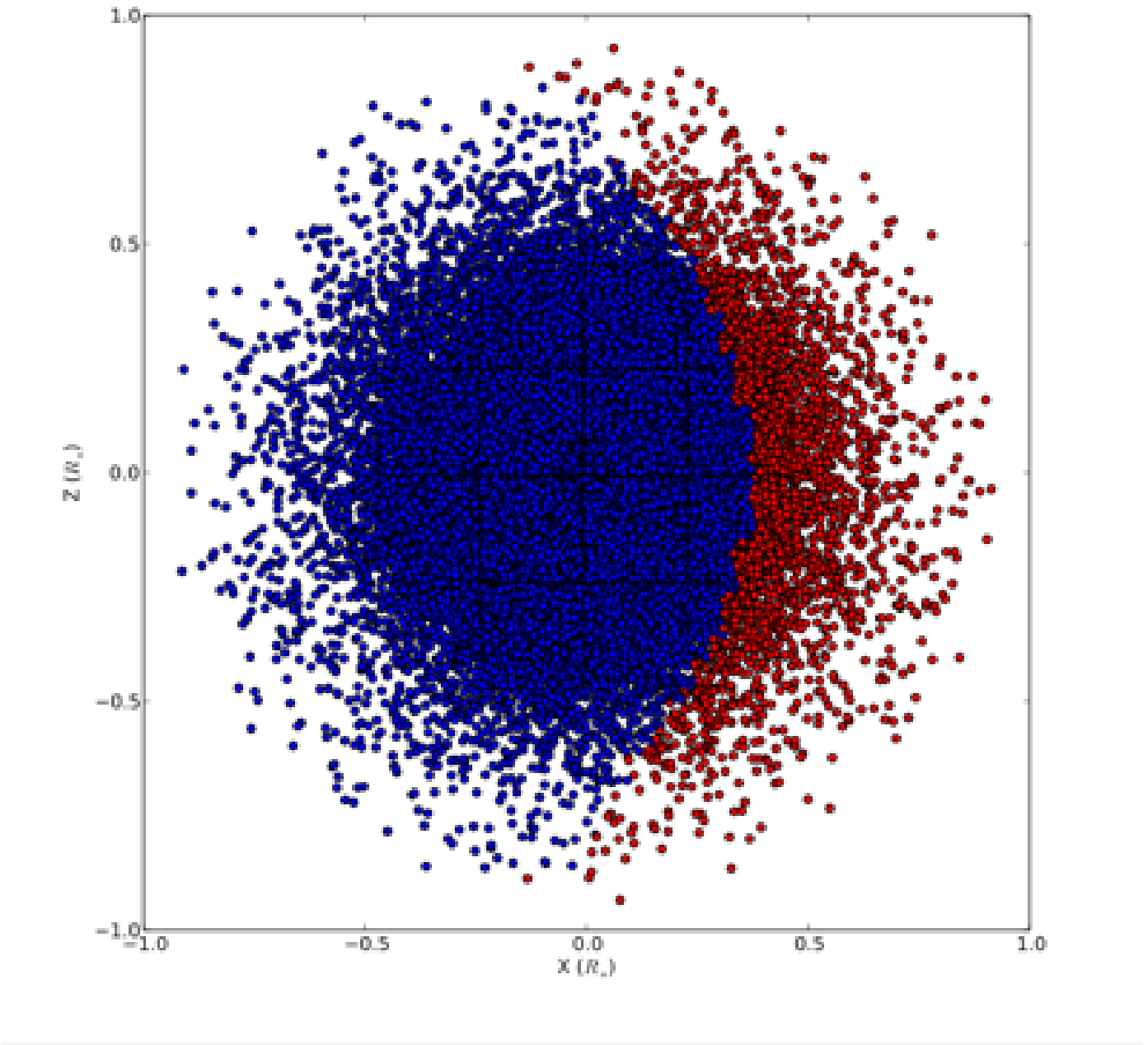}
\plotone{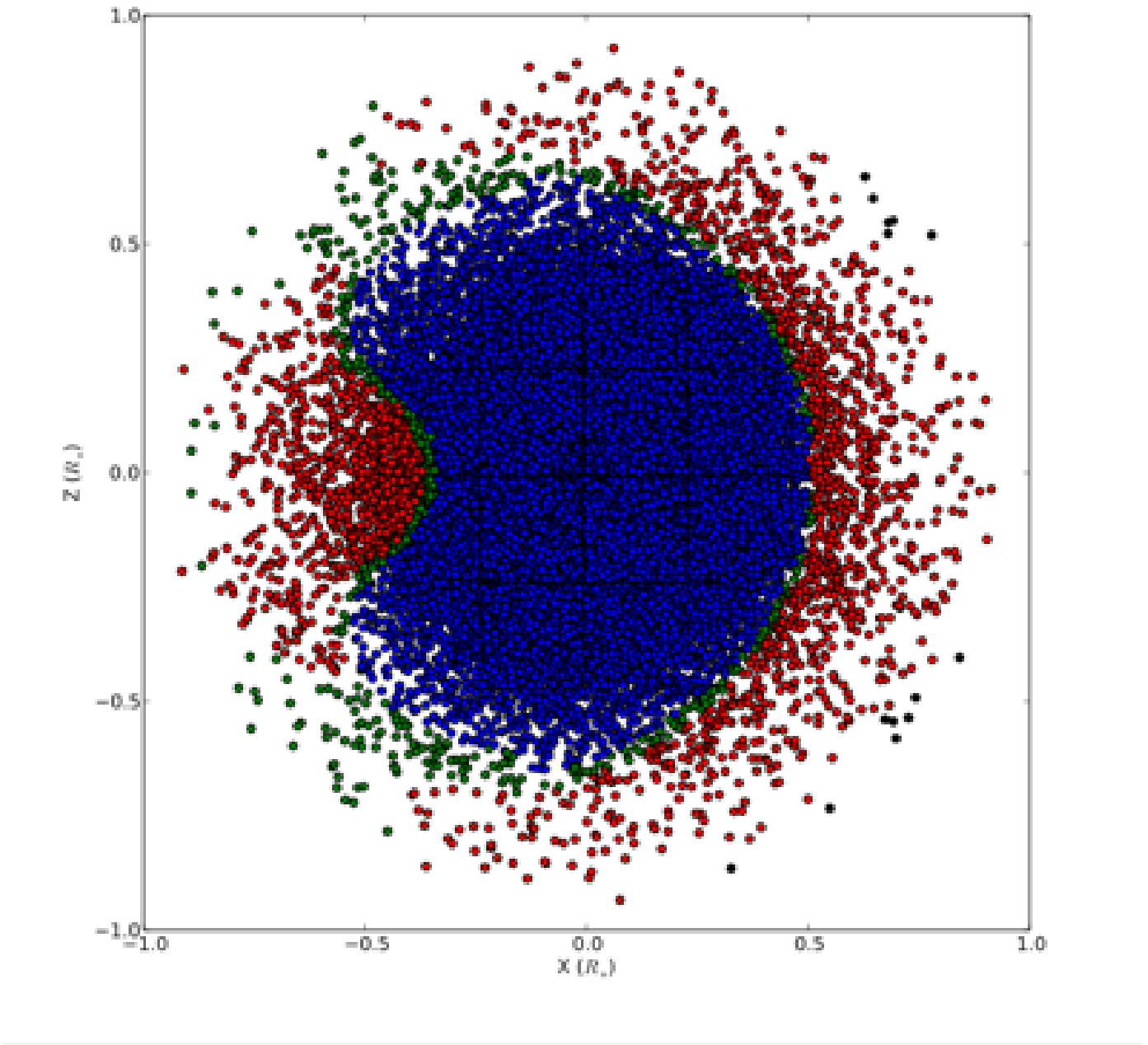}
\caption{\label{part_ms} Initial position of passive particles in the $x-z$ plane.
Particles are plotted within a slice of thickness equal to three zones. 
The axes are scaled to the radius of the companion star ($R_{\rm MS}=5.51 \times 10^{5}$~km). 
Colors represent the gravitational binding status at $t=144$~sec (upper) and $t=1060$~sec (lower) for the case MS-r (see Table~\ref{tabAllruns}): 
blue for bound; red for ablated; green for stripped; 
and black for particles already gone from the simulation box at that time. 
The black grids between particles reveal some numerical artifacts in particle initialization, but this artifact does not alter the dynamics of the fluid or the tracer particles. A detailed explanation and calibration are given in Section~\ref{sec_particles}.}
\end{figure}

By including the orbital motion and spin, the MS-WD scenario loses $16\%$ more mass for the MS-r run and $4\%$ for the MS-5 run.  The additional mass 
lost may not be accurate for runs with large separations: the mass loss decreases dramatically at larger separations.  Other 
uncertainties associated with the spatial resolution (or the gravitational 
potential in the binary system) may be comparable to the percentage of mass difference in runs characterized 
by large separations. 
In general the orbital motion should not change the binding energy significantly if the binary separation is larger than the Roche separation. 
In addition, the orbital speed is much smaller than the SN ejecta speed. 
Thus, orbital motion should not have much effect on the impact of the explosion, instead simply providing a constant velocity for the companion star. 
However, the spin of the companion may have a stronger influence. 
Adding rotational energy to the companion star should make its envelope's binding energy lower.
This addition may explain the extra mass lost in our simulations that include spin.
Figure~\ref{rotEner} shows the distribution of the ratio of rotational energy density and binding energy density in the $x-z$ plane.  
The rotational energy density in the back and front region of the companion star is $\sim 3\%$ of the binding energy density at $R\sim 0.8 R_*$.
When compared with the unbound region in Figure~\ref{part_ms}, there is evidence of some 
overlap of the unbound mass with the high rotational-energy density region. 
Therefore, the spin of the companion should broaden the unbound mass region.   

The simulation data from \cite{2000ApJS..128..615M} and \cite{2008A&A...489..943P} are also plotted and compared in Figure~\ref{fubSep}. 
Although the MS companion model in \cite{2000ApJS..128..615M} is different from that used here, 
the general behavior of their results is consistent with ours. 
\cite{2008A&A...489..943P} have a MS star model that resembles ours but is slightly larger.
The power-law relation for the final unbound mass versus separation provided by \cite{2008A&A...489..943P} also is consistent with our MS-WD scenario.

\cite{2007ApJ...670.1275L} studied the spectra of two SNe Ia (SN~2005am and SN~2005cf) in order to search for the H$\alpha$ emission that may emerge from the unbound hydrogen-rich companion mass.
He relied on the data from the two SNe and coupled the modeling results of \cite{2005A&A...443..649M} and the final unbound mass in \cite{2000ApJS..128..615M} and \cite{2007PASJ...59..835M}, 
providing an estimated upper limit of 0.02~$M_\odot$ for the amount of solar abundance material that may have remained undetected. 
In contrast, the mass loss for our cases in RLOF are all greater than these limits, except for the He-WDc scenario.
However, the limits should depend on the actual evolutionary stage of the companion candidate.
If the upper limit is correct, SN~2005am and SN~2005cf may result from a more compact binary companion or from the DDS.

Similarly, the kick velocity versus binary separation can also be described by a power-law relation,
\begin{equation}
\delta v_{\rm kick} = C_{\rm kick} a^{m_{\rm kick}} M_\odot, \label{eqkick}
\end{equation}
where $m_{\rm kick}$ is the power-law index, and $c_{\rm kick}$ is a constant depending only on the companion model (see Figure \ref{vkickSep}).
The power-law indices and constants for the final unbound mass and kick velocity are summarized in Table~\ref{tab3}.
For the RG-WD scenario, because the simulation domain is of a very large scale, the error in kick velocity is comparable to the value we deduced.
Therefore, we can only provide an upper limit on the kick velocity of about 
$40$~km/sec.

If we include the orbital motion and spin, they only produce about a $2\%$ difference in the kick velocity, which suggests that the orbital motion and spin are unimportant to the momentum transfer between ejecta and companion.

Although our final unbound mass is consistent with \cite{2008A&A...489..943P}, the kick velocity we determined is lower than the kick in \cite{2008A&A...489..943P}. 
This result may be due to the lack of a reverse shock in the SPH simulations: the reverse shock carries some momentum away from the SN ejecta, consequently making the companion kick smaller.

\subsection{Stripped and Ablated Mass \label{sec_particles}}

\begin{figure}
\plotone{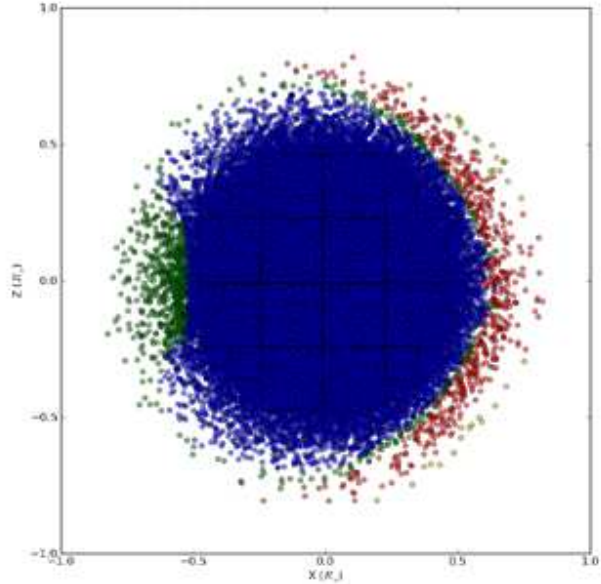}
\caption{\label{part_he} Similar to Figure~\ref{part_ms}, but for case He-r at $t=265$ second. }
\end{figure}
\begin{figure}
\epsscale{0.9}
\plotone{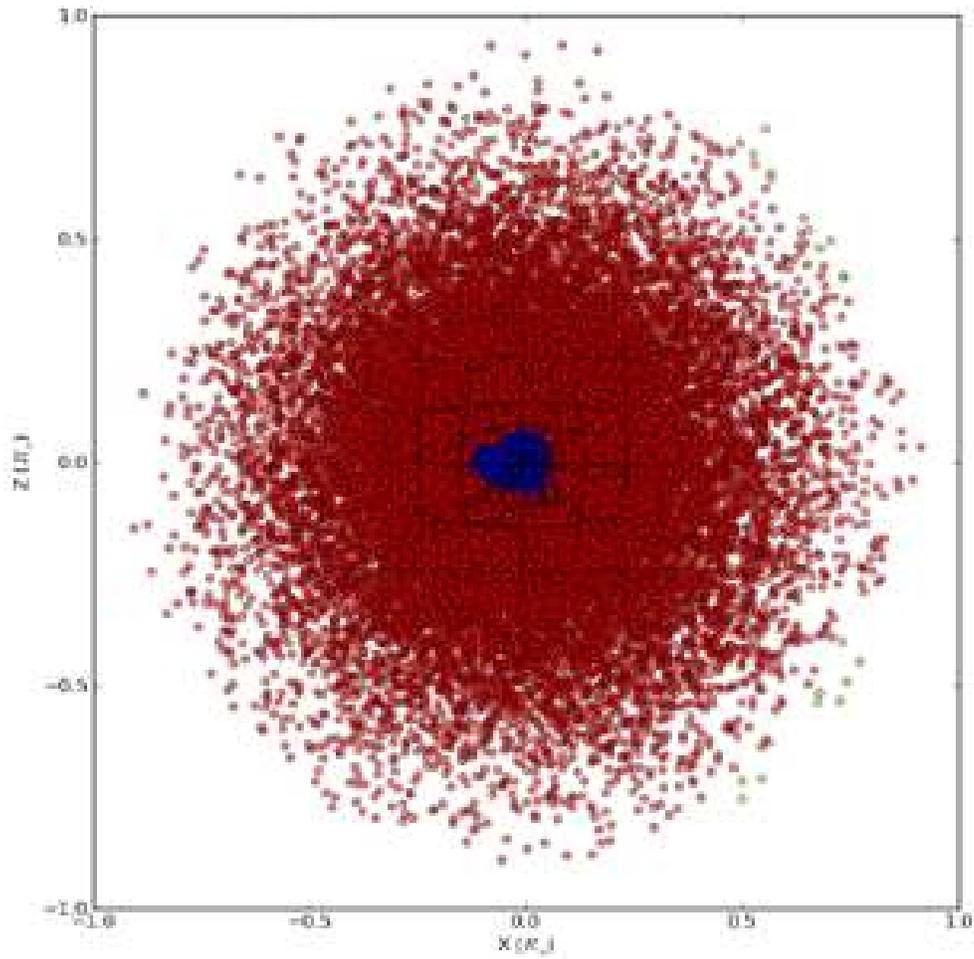}
\caption{\label{part_rg} Similar to Figure~\ref{part_ms}, but for case RG-r at $t=16.0$ hours. }
\end{figure}

The unbound mass could be either due to ablation (heating) or to stripping (momentum transfer), depending on the energy of the SN ejecta and the binding energy of the companion\citep{1975ApJ...200..145W}.
Quantitatively determining the ablated and stripped mass requires a method to trace the fluid elements in Lagrangian coordinates.
This method can be implemented in FLASH using passive particles.
We create an additional particle property for each passive particle and then label its gravitational-binding status as bound, ablated, or stripped.
$10^6$ passive particles are randomly distributed, with particle density proportional to the gas density. 
All passive particles are initialized as bound and then the gravitational-binding status in each timestep is recalculated.
Particles in zones with negative total energy ($E_{\rm tot}=E_{\rm grav} + E_{\rm kin} + E_{\rm thermal}$) are considered to be bound; otherwise, we mark passive particles in zones as ablated (stripped) if the thermal energy is larger (smaller) than the kinetic energy.
Ablated and stripped particles possibly become bound again in the subsequent 
timesteps, but they are not allowed to switch from one state to the other. 

Figure~\ref{part_ms} illustrates the initial distribution of passive particles colored by their gravitational-binding status in the $x-z$~plane. 
At $144$~seconds (upper panel), particles on the front side of the companion are all ablated as a result of the initial impact. 
The MS star companion is compressed and ablated to the initial radius of $\sim 0.4 R_*$.  
This result corresponds to the first peak of the unbound mass ($\sim0.2 M_\odot$) in comparison to the simulation time in Figure~\ref{convergence3d}.
Next, the reverse shock, traveling in the positive $x-$~direction, thins the unbound layer to $\sim 0.5 R_*$ and decreases the unbound mass. 
Following this decrease, the companion is surrounded by the ejecta while mass around the surface and in the tail region becomes unbound again (lower panel at 1060~sec), eventually increasing to $0.21 M_\odot$ at the end of the simulation (Figure~\ref{convergence3d}). 
This non-monotonic behavior of changing unbound mass in time also has been seen by \cite{2008A&A...489..943P} in their SPH simulations. 
They, however, do not examine the unbinding process at early stage and their non-monotonic behavior is relatively weak.
Therefore, they deemed it to be a numerical artifact in their convergence study.
However, it is difficult to definitively show that the non-monotonic behavior is numerical in origin.
 
In contrast to the MS-WD scenario, Figure~\ref{part_he} and Figure~\ref{part_rg} reveal similar distributions for the He-WD scenario and the RG-WD scenario.
The He-WD scenario's behavior resembles the MS-WD scenario, but mass is only ablated in the leading surface region.
More stripped mass in the tail region results from the high turbulence around the He star.
The dramatic impact and turbulence make momentum transport more efficient.
In the RG-WD scenario, almost all the envelope is ablated during the initial impact; only a 
small amount of mass ($\sim 0.007 M_\odot $) is stripped.
Later on, some mass around the core region becomes bound again after the RG is compressed.
After the impact, the bound mass slowly increases because of fallback.
 
The amount of ablated and stripped mass can be calculated by counting the ablated and stripped particles. 
Non-interacting passive particles do not affect the accuracy of the subsequent 
time evolution because passive particles simply move in Lagrangian coordinates without interacting with fluids.
However, the initialization of passive particles in our setup was found after our simulations to have some numerical artifact related to block boundaries.
This grid-shaped artifact can be seen in Figure~\ref{part_ms}. 
Therefore, a weighting factor for the ablated/stripped mass due to each particle is necessary when calculating the total particle mass.

We calculate the particle weights by evaluating the difference between particle density and gas density in each zone for the initial conditions. 
Thus, the particles in any given zone have the same weight.
However, in low-density regions, some zones may not contain any particles; 
this lack may lead to another uncertainty, underestimating the total particle mass.
Thus, we add this missing mass by assuming that the stripped mass to ablated mass ratio in these zones is the same as for their neighbors.  
After this correction, we are able to calculate the stripped mass and ablated
mass for each timestep.

The stripped mass to ablated mass ratios for all the runs are summarized in Table~\ref{tabAllruns}. 
These values are calculated from the unbound particles at a time after the initial impact and before the unbound particles leave the simulation box.
It is found that the unbound mass is mainly due to ablation in all the runs. 
The stripped mass to ablated mass ratio is about $0.4-0.7$ for the MS-WD scenario, $0.5-0.8$ for the He-WD scenario, and less than $0.2$ for the RG-WD scenario. 
This result contradicts our previous results using the mixing of SN ejecta at late times in  2D simulations \citep{2010ApJ...715...78P} (tracer particles are not supported in 2D).
The ablation happens at the very beginning of the initial impact, but the mixing of ejecta occurs later, based on our analysis using passive particles.
Thus, the mass treated as stripped in the previous study may already have been ablated during the initial impact.
In addition, ablation was previously ignored in most previous analytical or semi-analytical work 
\citep{2007PASJ...59..835M}, resulting in an underestimate of the impact of SN ejecta on the binary companion.

\subsection{Hole in the Supernova Remnant}

\begin{figure}
\plotone{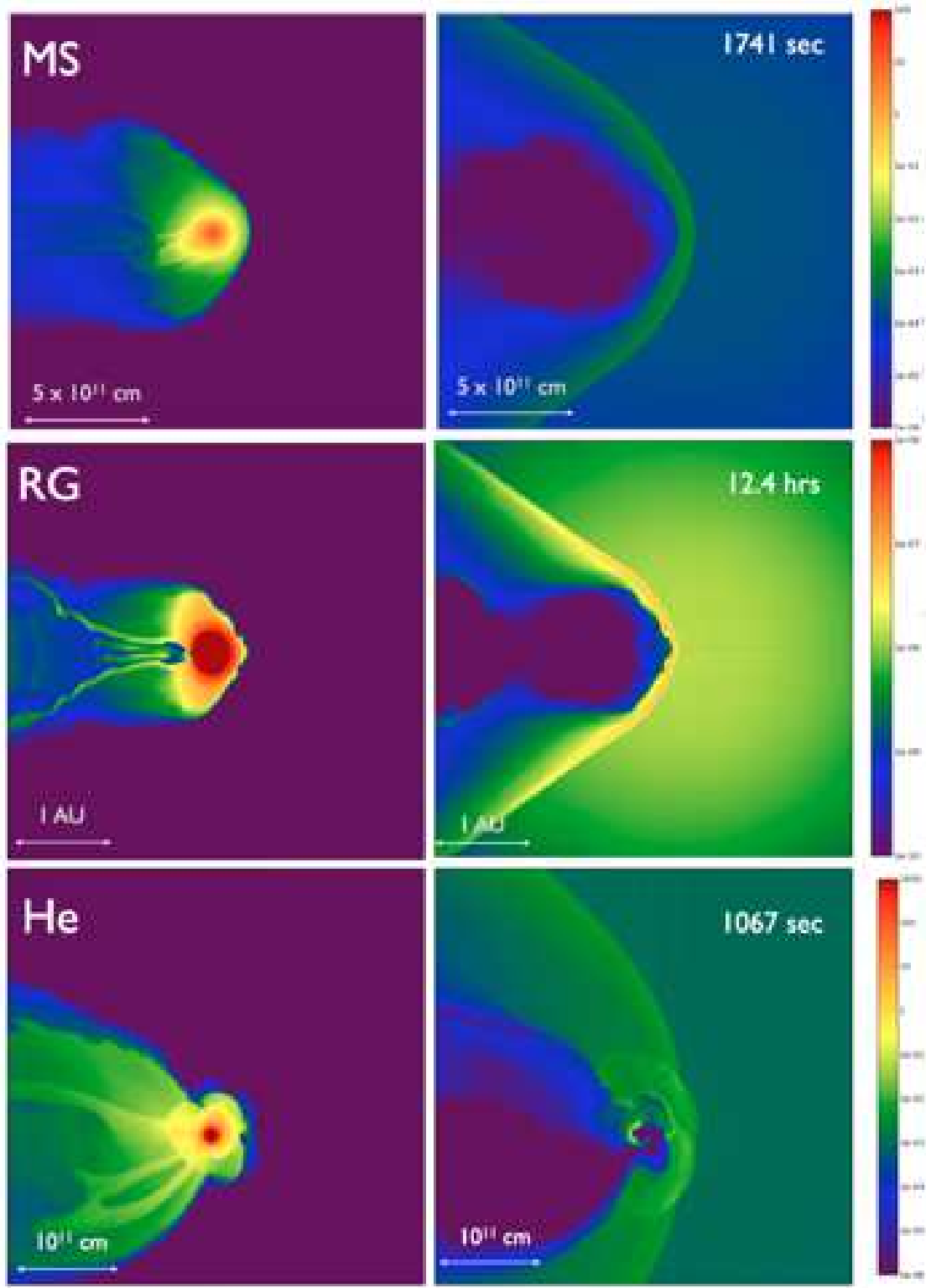}
\caption{\label{dens_cp_sn} Companion and SN gas density in the orbital plane for different scenarios. The left panel displays the density of the companion material ($^1H+^4He+^{12}C$) and the right panel shows the density of SN material ($^{56}Ni$). Companion model and initial setup from the top row to the lowest row are MS-r, RG-r, He-r cases in Table~\ref{tabAllruns}.}
\end{figure}

\begin{figure}
\plotone{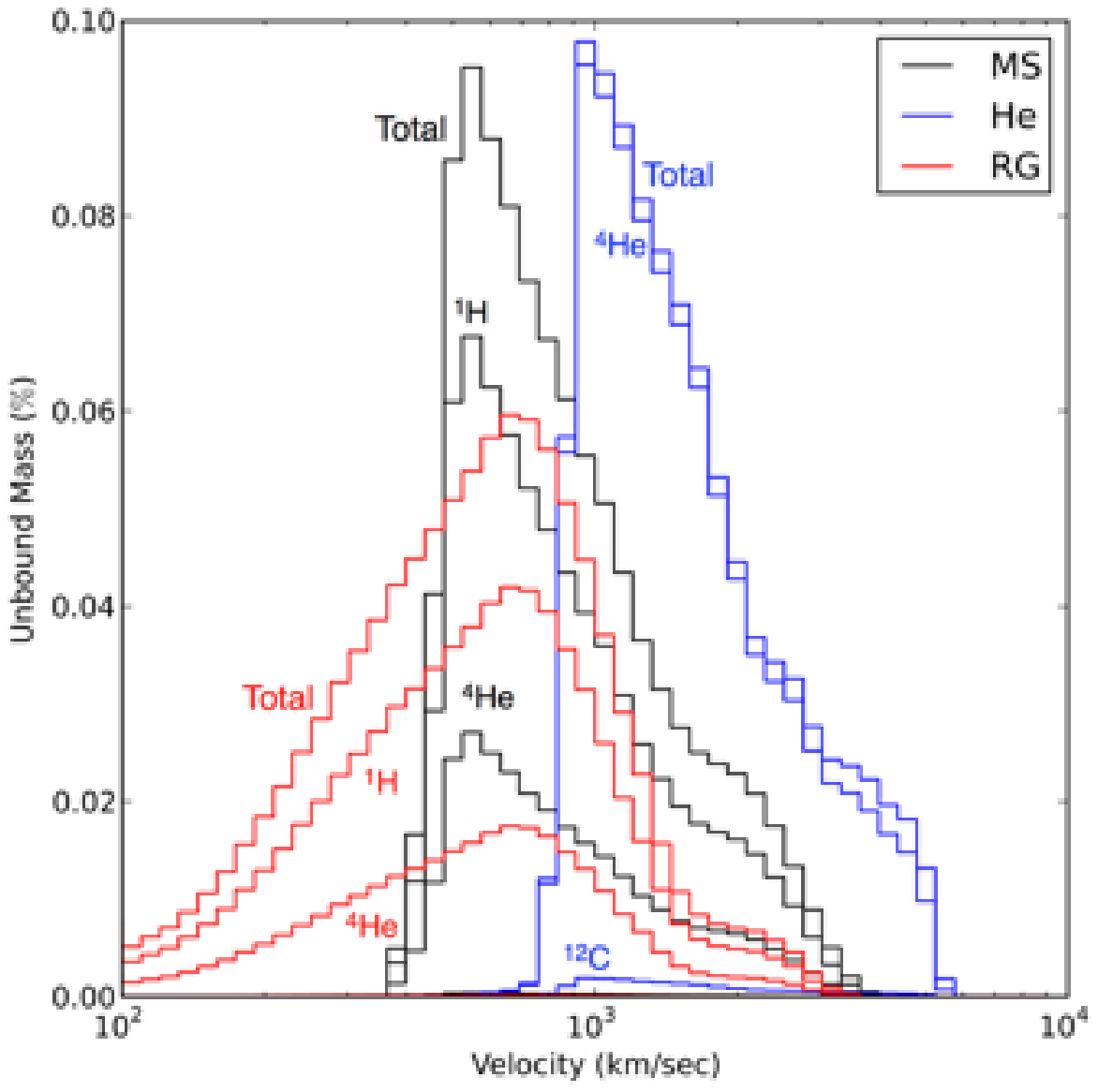}
\caption{\label{ubmass_vel} Speed distribution of unbound mass and compositions for the MS-r, RG-r, He-r cases (Table~\ref{tabAllruns}).  
Each color represents the speed distribution for a particular run: 
black represents the case MS-r; red, case RG-r; and blue, case He-r.}
\end{figure}

In the SDS, the impact of SN ejecta not only can affect the companion star, but it also can affect the shape of the ejecta: a cone-shaped hole shadowed by the companion will break the symmetry of the ejecta. 
Observationally, spectropolarimetry studies reveal a continuum polarization of about $0.2\% -0.7\%$ for normal SNe~Ia and $0.3\%-0.8\%$ for subluminous SNe~Ia \citep{2001ApJ...556..302H, 2008ARA&A..46..433W}, implying that the asymmetry in SN~Ia explosions is small but real.
Recent theoretical and numerical studies suggest that the intrinsic variation in SN~Ia light curves potentially results from viewing asymmetric explosions from different angles \citep{2004ApJ...610..876K, 2010ApJ...708.1025K, 2010Natur.466...82M, 2011MNRAS.417.1280B}. 
Furthermore, \cite{2012ApJ...745...75G} studied the interaction of the hole, SN material, 
and ambient medium using SPH simulations. They conclude that the hole could remain open in the SNR for hundreds of years, suggesting the hole could affect its structure 
and evolution.

Figure~\ref{dens_cp_sn} shows the companion-material and SN-material density distribution in the orbital plane for the MS-r, RG-r, and He-r cases. 
The companion material is confined mainly in the bow shock; the opening angle of the hole is about $40^\circ - 45^\circ$ for the MS-r case, $45^\circ-50^\circ$ for the RG-r case, and $30^\circ - 35^\circ$ for the He-r case. 
The MS star and RG companions have a similar opening angle, but the high-speed tail shock in the RG companion has the largest opening angle due to having the lowest ejecta Mach number.  
The high orbital speed and strong turbulence in the He-r case distort and mix the hole with more SN material than other companion models, creating a smaller hole with more fallback in the SNR. 

We may also ask how the unbound hydrogen-rich material is distributed in the SNR and how this material can be detected by observations.
Based on our simulations, the unbound mass is mostly confined in the hole. 
The velocity distribution of unbound mass is shown in Figure~\ref{ubmass_vel}. 
The peak velocity is $550$km/sec$^{-1}$ for the MS-WD, $955$km/sec$^{-1}$ for the He-WDc, and $660$km/sec$^{-1}$ for the RG-WD.
The RG companion has the largest velocity dispersion, but the peak velocity is slightly larger than the peak velocity for the MS star companion. 
This low outflow velocity ($< 1,000$km/s$^{-1}$) is much smaller than the ejecta speed ($\sim 8,000$km/s$^{-1}$), suggesting that the hydrogen could be hidden in the ejecta in the early-stage explosion, except when looking directly from the backside of the companion. 
However, it is still possible to detect it at late times when the expansion slows down and the ejecta becomes transparent.

\subsection{Nickel Contamination}

\begin{figure}
\plotone{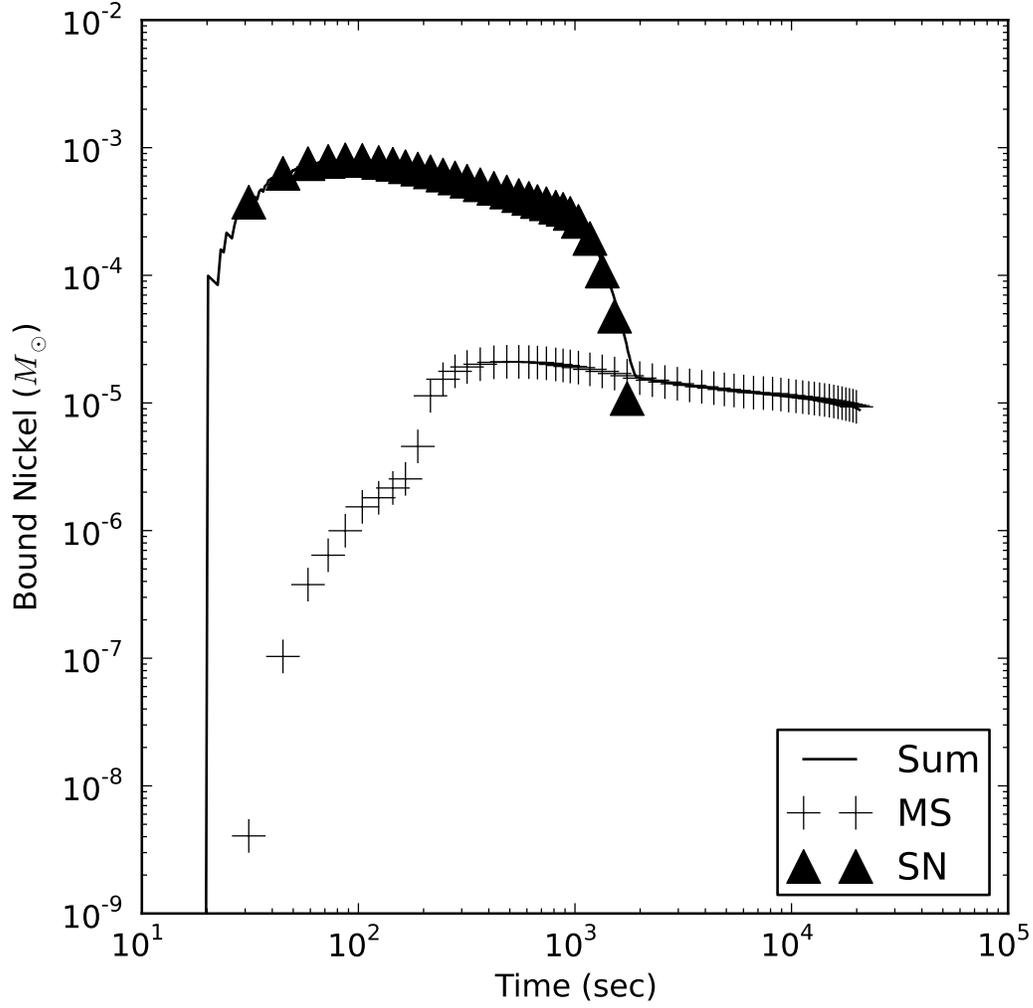}
\caption{\label{bniMS} Bound nickel around the MS star companion region and the SN region.
The plus sign indicates the companion region, and triangles indicate the SN region. 
Total bound nickel is plotted using the solid line.}
\end{figure}

The companion star could be contaminated by the SN ejecta in its envelope during the initial impact or as fallback at late times.
If the mixing of SN ejecta occurs mainly in the companion's envelope, the contamination provides one possibility for detecting iron abundance from the remnant star after the nickel radioactively decays.
Since we use nickel as a tracer for the SN~Ia ejecta in our simulations, we could estimate the nickel contamination in the remnant star in a SNR by calculating the amount of bound nickel after the SN~Ia explosion.
Following the same method as that used to determine the bound mass, we can calculate the amount of 
bound nickel by summing the ejecta material that cannot escape the gravitational potential.   
There are essentially two bound regions in the simulations: the region around the post-impact 
companion star and the region at the explosion center, separated by the reverse shock. 
The bound nickel in the SN region likely results from the fact that the some of the nickel is bound to the system in the initial setup due to the presence of the companion star. We assume that the bound nickel in the companion star more likely reveals the nickel contamination.

Figure~\ref{bniMS} illustrates the amount of bound nickel in these two regions for the case MS-r.
The bound nickel is dominated by the SN region at the beginning and then reaches a 
peak of $\sim 10^{-3} M_\odot$ at about $\sim 80$~sec. 
However, after the interaction between the reverse shock and the SN region, the bound nickel in the SN region becomes unbound and eventually disappears after around $2,000$~sec. 
The remaining bound nickel in the MS star companion is $\sim 10^{-5} M_\odot$. 
In general, increasing the initial binary separation leads to a greater 
contamination because a slower ejecta speed makes it more difficult for ejecta material to escape; a larger separation, however, leads to a smaller companion cross-section, thereby blocking less SN material.
Thus, a simple variation for the nickel contamination with the binary separation is not found. 
All the results are consistent to within an order of magnitude with a value of $\sim 10^{-5} M_\odot$.    
Orbital motion and spin only introduce a minimal effect.
This amount of nickel contamination is significantly less than the previous estimate by  \cite{2000ApJS..128..615M} for the HCV scenario (an upper limit of $10^{-3} M_\odot$) because the interaction of the reverse shock.

\begin{figure}
\plotone{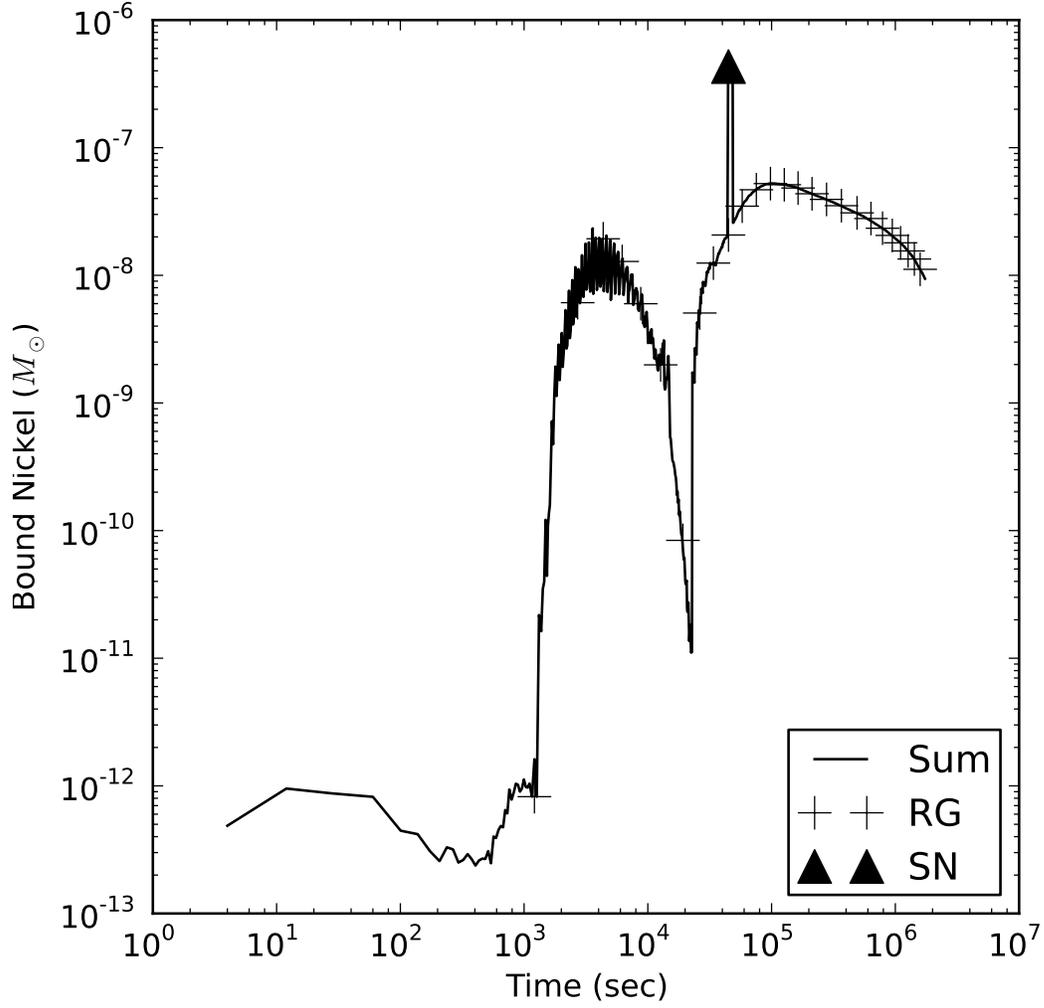}
\caption{\label{bniRG}Similar to Figure~\ref{bniMS}, but for RG binary companion. Note that there is only one triangle data point for SN region because the duration of this region is short. We have only one checkpoint file during that time.}
\end{figure}

Figure~\ref{bniRG} shows the nickel contamination for the RG companion, the RG-r case.
In this example, the bound nickel in the SN region only exists for a very short time at about $10^5$~sec after the explosion. 
There are two bumps of bound nickel around the RG region.
The first bump, at about $\sim 4,000$~sec, is related to the initial impact of the SN~Ia explosion.
Bound nickel reaches the first peak when the shock compresses the leading surface of the RG, and then it decreases when the reverse shock passes through it. 
The second bump occurs later at about $\sim 10^5$~sec in a region at the back side of the RG while the compressed RG relaxes.    
The nickel contamination for the case RG-r is about $10^{-8} \sim 5 \times 10^{-8} M_\odot$.

\begin{figure}
\plotone{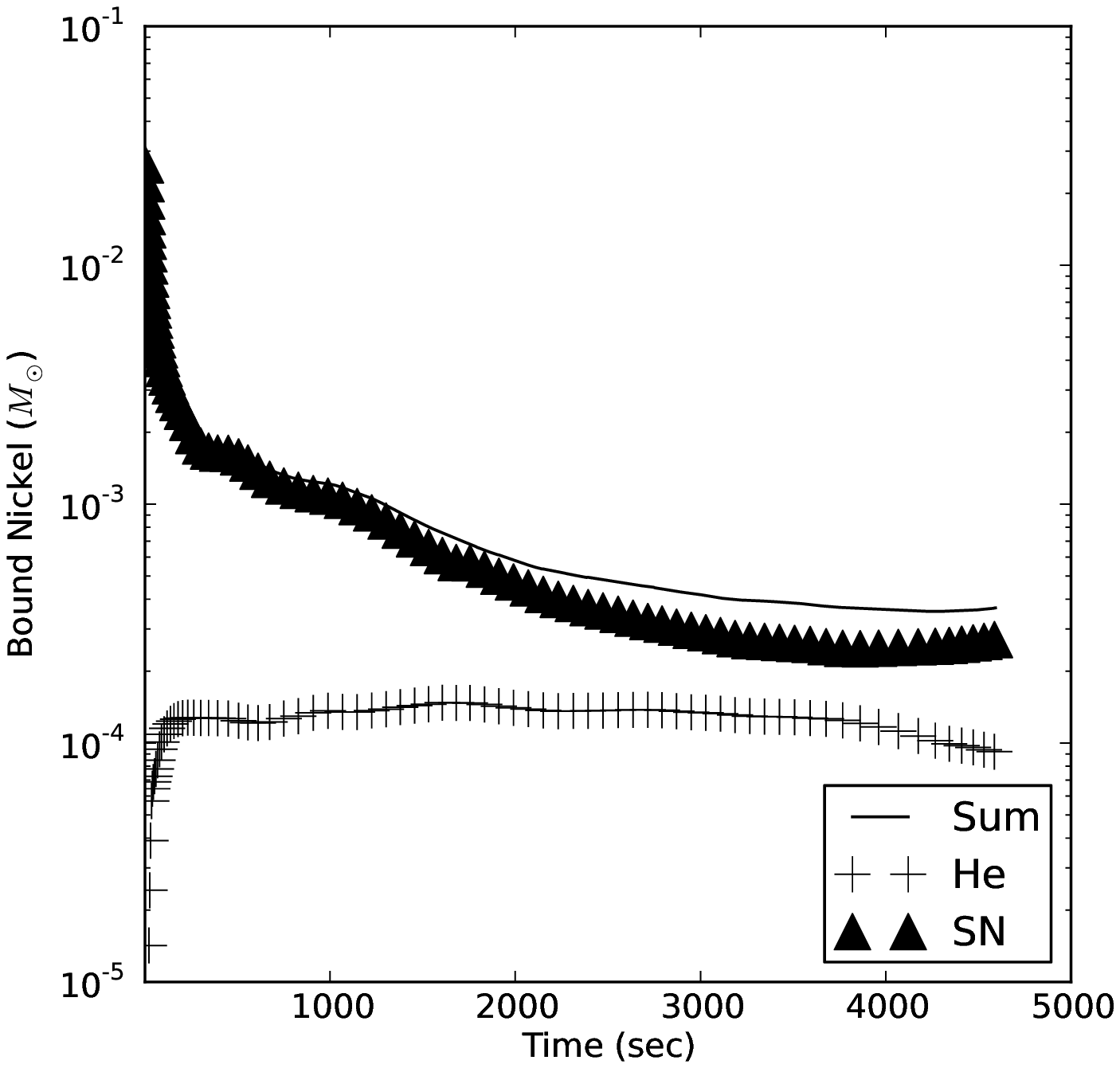}
\caption{\label{bniHe}Similar to Figure~\ref{bniMS}, but for helium star binary companion.}
\end{figure}

\begin{figure}
\plotone{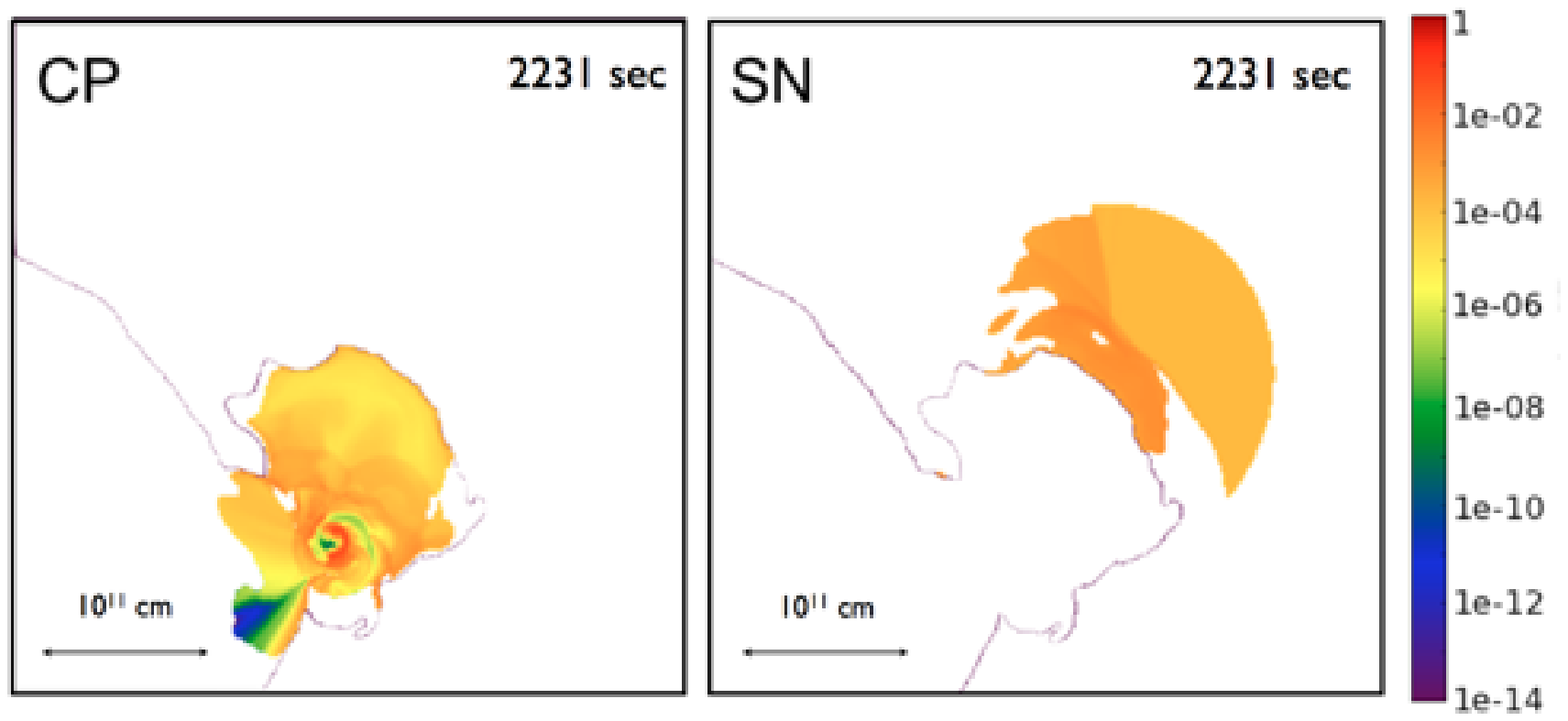}
\caption{\label{bniHe_dens} Bound nickel density distribution in the orbital plane around the helium star companion (left) and the SN region (right).
The black line describes the boundary between the companion material region and the SN material region. 
The color scale indicates the logarithm of the nickel density in g~cm$^{-3}$.}
\end{figure}

Similarly, Figure~\ref{bniHe} shows the nickel contamination for the helium star companion (case He-r).
The nickel contamination in the He star region stays around $\sim 10^{-4} M_\odot$ for the whole simulation, but in the SN region it decreases from $10^{-3} \sim 10^{-2} M_\odot$ to $\sim 3\times 10^{-4}$. 
Unlike the MS star cases, the bound nickel in the SN region does not disappear at the end of the simulation while interacting with the reverse shock.  
It is more likely that the bound nickel in the SN region will merge eventually with the He star region at late times, resulting in a total contamination of $\sim 4 \times 10^{-4} M_\odot$. 
Figure~\ref{bniHe_dens} shows the bound nickel density in the orbital plane at $2,231$~sec for these two regions; at that time, the two regions are already mixed with each other.

If the nickel contamination is restricted to the envelope of the companion star, we can estimate the nickel to hydrogen plus helium ratio by using the bound nickel and the envelope mass \citep{2010ApJ...715...78P}.
We define the envelope radius using the extreme in second derivative of the companion gas density with respect to the radius in the initial conditions. 
Thus, the envelope mass is the mass outside this envelope radius.
If we assume the unbound mass is entirely from the envelope (Figure~\ref{part_ms}), then the final envelope mass is the difference between the initial envelope mass and the final unbound mass in Table~\ref{tabAllruns}. 
However, we note that the envelope mass ($\sim 0.3 M_\odot \pm 0.1M_\odot$) we estimate for the MS star companion is close to the final unbound mass ($\sim 0.2 M_\odot$); thus for this case we only can provide an order of magnitude estimate.
The estimated upper limit for the ratio of nickel to hydrogen plus helium is about $10^{-4}$ for the MS-WD, $10^{-3}$ for the He-WDc, and $2 \times 10^{-5} - 10^{-6}$ for the RG-WD scenario.
The solar ratio from \cite{1989GeCoA..53..197A} is about $5 \times 10^{-4}$, which is similar to or slightly smaller than the results in our simulation values, suggesting a possible probe to identify the progenitor candidate in a SN~Ia SNR.

\subsection{The Remnant Companion Star}

\begin{figure}
\plotone{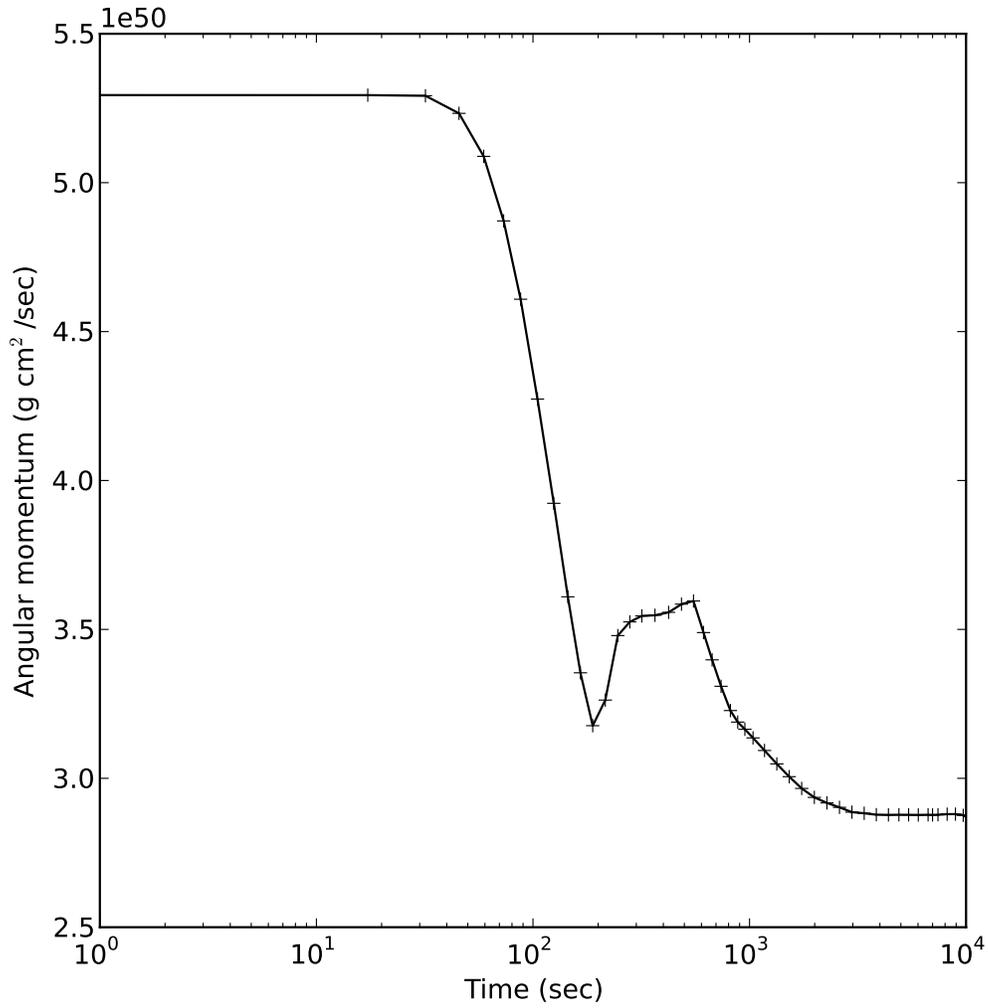}
\caption{\label{angular}Angular momentum versus simulation time for the MS-r case. The angular momentum is calculated in the companion center of mass frame and only considers the  bound companion gas. }
\end{figure}

At the end of our simulations, all the companion stars have survived the impact of the SN~Ia explosion.
The post-impact velocity of the remnant star will be the original orbital velocity plus the kick velocity.
In general, this velocity should be larger than the velocities of background stars.
In our simulations, the kick velocity contributes up to $45\%$ of the final net velocity for the MS-r case, $13 \%$ for the He-r case and $< 50 \%$ for the RG-r case.
Therefore if the binary systems in the SDS are in RLOF, then the final velocity including kick and orbital speed will provide information in the search for the progenitor star in the SNR.

\cite{2004Natur.431.1069R} observed stars near the remnant center in Tycho's SN. 
They found a star named Tycho G which has a higher radial velocity ($108 \pm 6$km~s$^{-1}$, in the local standard of rest) than other background stars at the corresponding distance ($20-40$ km~s$^{-1}$ at $2-4$~kpc), suggesting Tycho~G could be the non-degenerate remnant star resulting from the SDS. 
Tycho~G is a slightly evolved star with a stellar type G0-G2, 
mass $M\sim1 M_\odot$, radius $R\sim 1-3 R_\odot$, and effective 
temperature $T_{\rm eff}=5,900 \pm 100$~K  \citep{2009ApJ...691....1G}. Thus, if it is indeed a remnant, the progenitor star was likely 
a MS star. 
Although our simulation is not set up for Tycho's SN, the MS-r case resembles it.
 The final net velocity of our MS-r case is $277$km~s$^{-1}$, which covers the value of Tycho~G.  
If the remnant star in Tycho's SN has not changed significantly since the SN explosion,  
the property of our post-impact companion may provide further information for identifying the progenitor star of Tycho's SN.
The post-impact central density, temperature, and pressure in our simulation  
for the MS-WD model are $24$~g~cm$^{-3}$, $7.8\times 10^6$~K, 
and $2.7\times 10^{16}$~dyne~cm$^{-2}$, which are consistent with the 
values in \cite{2000ApJS..128..615M}. 

As \cite{2011ScChG..54.2296M} pointed out, the rotational velocity of the remnant star could be 
an important diagnostic for the Tycho G star as the companion candidate in Tycho's SN with the SDS.
In their model, the mass, space velocity, radius, luminosity, and effective temperature of the remnant star are consistent with observations by \cite{2004Natur.431.1069R} and \cite{2009ApJ...701.1665K}, but not the rotational velocity. 
The rotational velocity observed by \cite{2009ApJ...701.1665K} is only $7.5 \pm 2$~km~s$^{-1}$, but the predicted rotational velocity in \cite{2011ScChG..54.2296M} is 
$\sim 100$~km~s$^{-1}$.  However, \cite{2011ScChG..54.2296M}
ignore the impact of the SN ejecta on the companion star and assume the remnant star did not change at the onset of the SN~Ia explosion.  
Our simulation shows that about $18\%$ of the mass is lost in the MS-r case, 
suggesting that the angular momentum of the remnant star also should decrease after the impact.
    
Figure~\ref{angular} shows the change of angular momentum versus simulation time in the companion center of mass frame. 
After the impact, the MS companion loses $48\%$ of its initial angular momentum ($6\%$ for the He-r case and $\sim 99\%$ for the RG-r case), but only loses $18\%$ of the mass. 
If the angular momentum, $J = \alpha M_* R_*^2 \omega_*$, has a constant $\alpha$ after the impact of SN explosion, then $\alpha$ can be calculated using the angular momentum from the initial conditions ($\alpha= J_0/(M_{*} R_{*}^2 \omega_{*})=0.251$).
However, the post-impact radius of the companion is not straightforward to define, 
because the remnant companion star is not in hydrostatic equilibrium and the location 
of the photosphere is unclear. 
We can estimate the equilibrium radius of the remnant star in the case MS-r using the 
virial theorem, $R_{\rm remnant} \sim G M^2/ E_{\rm int} = 1.6 \times 10^{11}$~cm$= 2.4 R_\odot$.
Therefore, the post-impact rotational velocity can be calculated from the post-impact $R'$, $M'$ and $J'$ to obtain $v_r' \sim 37$ km~s$^{-1}$, which is only $23\%$ of the initial rotational velocity ($v_{r,0} = R_* \omega_* = 164.4$ km~s$^{-1}$). 
However, the virial theorem and constant $\alpha$ assume hydrostatic equilibrium, and Tycho's SN is only $439$~yr old. The real $R'$ should be larger, causing a lower rotational velocity for the post-impact companion star.
This result provides additional support for the interpretation of Tycho G as the candidate for the progenitor star in Tycho's SN with the SDS.

In contrast, the luminosity and surface temperature of the He star in the He-WD scenario are about $1.3\times 10^2 L_\odot$ and $4.4\times 10^4$~K, which are too high to be consistent with the candidate progenitor for the Tycho's SN.  
This would also be the case for the RG-WD scenario since the luminosity of the remnant is expected to be comparable to its RG progenitor, which must be sufficiently luminous.  
This is required in the SDS model in order that the mass transfer rate prior to the SN 
explosion be greater than $\sim 10^{-7} M_\odot yr^{-1}$ such that the CO WD accumulate sufficient matter to reach the Chandrasekhar limit.


\section{Conclusions}

We have investigated the impact of SN~Ia ejecta on companion stars in the single-degenerate scenario via three-dimensional hydrodynamical simulations.
We studied possible binary companion models, including a MS star, a RG and a He star, and considered the effects of asymmetry introduced by orbital motion and spin.
A detailed setup of the sub-grid SN~Ia explosion using the W7 model in \cite{1984ApJ...286..644N} also is described. 
It is found that the orbital motion and spin lead to $\sim 16\%$ more unbound mass in the MS star companion channel but do not significantly affect the kick velocity.
Furthermore, the orbital motion and spin play an important role in determining the morphology of the SNR.
A power-law relation between the unbound mass and initial binary separation is found for 
all companion channels and is consistent with previous studies. 
Similarly, the kick velocity can be fitted by a power-law for the MS and He star binary companions. 
For the RG companion, we can only report a $40$~km~s$^{-1}$ kick as an upper limit due to numerical uncertainty.   
By using the technique of passive particles, we find that the unbound mass is dominated by ablation instead of stripping.
This result is in conflict with previous understanding, in which ablation was 
ignored in previous analytical studies. In addition, a hole shadowed by the ejecta is found to break the symmetry of the SNR.
The amount of nickel contamination of the companion star is found to be $\sim 10^{-5}M_\odot$ for the MS star companion, $\sim 10^{-8}M_\odot$ for the RG companion, and $\sim 10^{-4}M_\odot$ for the He star companion. 
The corresponding nickel/iron to hydrogen plus helium abundance ratio may be useful for identifying the progenitor candidate in SN~Ia remnants in future observations.
We also find that the post-impact companion star loses about half of its initial angular 
momentum for the MS-WD scenario with the rotational velocity decreasing to $23\%$ 
of its initial rotational velocity, providing added support for the SDS model 
for the Tycho SN.


\acknowledgments
The simulations presented here were carried out using the NSF Teragrid's Ranger system at the 
Texas Advanced Computing Center under allocation TG-AST040034N.  FLASH was developed largely by 
the DOE-supported ASC/Alliances Center for Astrophysical Thermonuclear Flashes at the University of 
Chicago.  This work was partially supported by NSF AST-0703950 to Northwestern University and by the Computational Science and Engineering (CSE) fellowship at the University of Illinois at Urbana-Champaign.
Analysis and visualization of simulation data were completed using the analysis toolkit {\tt yt} 
\citep{2011ApJS..192....9T}.

\bibliography{snb3dRef}


\end{document}